\documentclass[journal,draftcls,twocolumn,10pt,twoside]{IEEEtranTCOM}
\usepackage{parskip}

\usepackage{mathtools, cuted}
\usepackage{lipsum, color}
\usepackage{setspace}
\usepackage{lipsum}
\usepackage{amsmath}
\usepackage{amsmath,amsthm,amssymb}
\usepackage{color,soul}
\usepackage{amsfonts}
\usepackage{epsfig}
\usepackage{cite}
\usepackage{subfigure}
\usepackage{multirow}
\usepackage{rotating}
\usepackage{graphicx}
\usepackage{tabularx}
\usepackage{array}
\usepackage{amsfonts}
\usepackage{graphicx}  
\usepackage{graphics}  
\usepackage{epsfig}
\usepackage[all]{xy}
\usepackage{tabularx}
\usepackage{amsmath}
\usepackage{amsfonts}
\usepackage{epstopdf}
\usepackage{cite}
\usepackage{subfigure}
\usepackage{multirow}
\usepackage{rotating}
\usepackage{graphicx}
\usepackage{tabularx}
\usepackage{array}
\usepackage{setspace}
\usepackage{morefloats}
\usepackage{float}
\usepackage[Euler]{upgreek}
\usepackage{mathtools}
\usepackage[mathscr]{eucal}

\makeatletter
\newcommand*{\rom}[1]{\expandafter\@slowromancap\romannumeral #1@}
\makeatother
\usepackage{amsmath}
\usepackage{algorithm}
\usepackage{algpseudocode}
\usepackage{float}
\usepackage{cite}
\usepackage{algorithm,algpseudocode}
\usepackage[margin=15mm]{geometry}
\usepackage{multicol}
\usepackage{float}
\usepackage{lipsum}

\makeatletter
\newcommand{\algmargin}{\the\ALG@thistlm}
\makeatother
\newlength{\whilewidth}
\algdef{SE}[parWHILE]{parWhile}{EndparWhile}[1]
{\parbox[t]{\dimexpr\linewidth-\algmargin}{%
		\hangindent\whilewidth\strut\algorithmicwhile\ #1\ \algorithmicdo\strut}}{\algorithmicend\ \algorithmicwhile}%
\algnewcommand{\parState}[1]{\State%
	\parbox[t]{\dimexpr\linewidth-\algmargin}{\strut #1\strut}}
\makeatletter
\def \BState{\State \hskip - \ALG@thistlm}
\makeatother
\usepackage[table]{xcolor}

\usepackage{amsmath, amsthm, amssymb}
\usepackage{amsthm}
\usepackage{etoolbox}
\usepackage{amsmath,amsthm}
\usepackage{lipsum}
\usepackage{amsthm}
\usepackage{amsfonts}
\usepackage{booktabs}
\usepackage{graphicx}
\usepackage{epstopdf}
\usepackage{epsfig}
\usepackage{amssymb}
\usepackage{amsmath}
\usepackage{cite}
\usepackage{color,soul}
\usepackage{subfigure}
\usepackage{multirow}
\usepackage{rotating}
\usepackage{tabularx}
\usepackage{array}
\usepackage{dblfloatfix}
\usepackage{blindtext}
\usepackage[normalem]{ulem} 

\newtheorem{definition}{Definition}
\begin{document}
	\title{Age of Information Aware VNF Scheduling in Industrial IoT Using Deep Reinforcement Learning
		\author{Mohammad~Akbari, Mohammad Reza~Abedi, Roghayeh~Joda, Member, IEEE, Mohsen~Pourghasemian, Nader~Mokari, Senior Member, IEEE, and Melike~Erol-Kantarci, Senior Member, IEEE}
		\thanks{Mohammad Akbari is with Communication Department, ICT Research Institute, Tehran, Iran (e-mail: m.akbari@itrc.ac.ir)}\thanks{Mohammad Reza Abedi is with the Department of Electrical Engineering, Tarbiat Modares
			University, Tehran, Iran (e-mail: Mohammadreza$\_$abedi@modares.ac.ir)}
		\thanks{Roghayeh Joda is with Communication Department, ICT Research Institute, Tehran, Iran (e-mail: r.joda@itrc.ac.ir). She is currently a visiting researcher at  the School of Electrical Engineering and Computer Science, University of Ottawa, Ottawa K1N 6N5, Canada (e-mail: rjoda@uottawa.ca)}
		\thanks{Mohsen Pourghasemian is Research Assistant at the Department of Electrical Engineering, Tarbiat Modares
			University, Tehran, Iran.}\thanks{N. Mokari is with the Department of Electrical Engineering, Tarbiat Modares	University, Tehran, Iran (e-mail: nader.mokari@modares.ac.ir). The work of Nader Mokari, Mohammad Reza Abedi, and
			Mohsen Pourghasemian  was supported by the Iran National Science
			Foundation under Grant No. 98025206.}\thanks{Melike Erol-Kantarci is with the School of Electrical Engineering and Computer Science, University of Ottawa, Ottawa K1N 6N5, Canada (e-mail: melike.erolkantarci@uottawa.ca)}}
	\maketitle
	\markboth{IEEE Journal on Selected Areas in Communication,~Vol.~XX, No.~XX, XXX~2021}
	{}
	\maketitle
	\begin{abstract}
		In delay-sensitive industrial internet of things (IIoT) applications, the age of information (AoI) is employed to characterize the freshness of information. Meanwhile, the emerging network function virtualization provides flexibility and agility for service providers to deliver a given network service using a sequence of virtual network functions (VNFs). However, suitable VNF placement and scheduling in these schemes is NP-hard and finding a globally optimal solution by traditional approaches is complex. Recently, deep reinforcement learning (DRL) has appeared as a viable way to solve such problems. In this paper, we first utilize single agent low-complex compound action actor-critic RL to cover both discrete and continuous actions and jointly minimize VNF cost and AoI in terms of network resources under end-to-end Quality of Service constraints. To surmount the single-agent capacity limitation for learning, we then extend our solution to a multi-agent DRL scheme in which agents collaborate with each other. Simulation results demonstrate that single-agent schemes significantly outperform the greedy algorithm in terms of average network cost and AoI. Moreover, multi-agent solution decreases the average cost by dividing the tasks between the agents. However, it needs more iterations to be learned due to the requirement on the agents' collaboration. 
	\end{abstract}
	\section{Introduction}\label{Introduction}
	\subsection{Background and Motivation}
	\subsubsection{Industrial Internet of Things} Future applications of Industrial Internet of things (IIoT) such as autonomous cars, virtual reality, and traffic control will increasingly focus on the exchange of delay-sensitive information for monitoring and control. In the application of virtual reality, the user's position and service information must be exchanged immediately between the server and users \cite{gubbi2013internet,zhou2019joint,al2015internet}. Therefore, in such applications, information must be kept fresh and outdated information is not valuable. 
	\subsubsection{Age of Information}
	Traditional metrics, such as packet delay and inter-delivery times used for real-time applications, are not sufficient to measure the freshness of information at the destination. Hence, the age of information (AoI) is considered as a criterion for evaluating the freshness of information received at the destination \cite{zhou2019joint,zhou2019minimum}. AoI increases linearly over time until next fresh packet is arrived, at which point AoI takes the value of the end-to-end packet delay \cite{sun2019age}. AoI can be used to characterize the information freshness since it is a powerful metric to capture the randomness of state updates. Meanwhile, the management of distributed, intelligent, and autonomous machines in a reliable and robust manner is the use of Network Function Virtualization (NFV). The NFV virtualizes network services and abstracts them from any dedicated hardware. It enables rapid service provisioning and chaining in IIoT \cite{8402355}.
	\subsubsection{Network Function Virtualization} NFV is an emerging network architecture which gives flexibility and agility to networks by realizing virtual network functions (VNFs) in software and adaptively placing virtualized services on physical resources. To provide an end-to-end service between two hosts, multiple VNFs might be used in sequence each of which deliver a specific network function. Suitable placements of VNFs and their routing over the available NFV infrastructure are important problems and must optimally be determined. Moreover, VNF placement should be dynamically adjusted to adopt to the network traffic and load changes over time \cite{pei2018efficiently}. NFV significantly improves scalability of networks, and at the same time enhances the robustness and agility in managing network components and makes them easy to deploy. In addition to these advantages of implementing network functions as VNFs, there is a tradeoff between flexible management and the imposed delay in placement and scheduling process of VNFs in emerging dense and complex networks composed of several VNFs. In IIoT applications, this issue is much more crucial \cite{9305697,8402355,8972306}. In our proposed IIoT NFV-enabled model, we address the robust and flexible management of very delay sensitive applications by distributing the functions of these applications into several virtualized servers. Accordingly, more delay will be imposed to the network by VNF placement and scheduling of the service functions. Therefore, AoI as a comprehensive metric for quantifying the freshness of information in an end-to-end manner, when it is formulated correctly to reflect all the components affecting information aging, can support the restrict delay requirements of IIoT applications in an efficient manner.
	\subsubsection{Machine Learning} 
	The VNF placement and routing can be expressed as a mathematical optimization
	problem with a constraint set that should be satisfied to meet the network infrastructure's restrictions and the service requirements. Nevertheless, such problems are NP-hard and finding globally optimal solutions is very difficult, especially in large scale networks \cite{marotta2017energy,liu2015improve}. Today, deep reinforcement learning (DRL) appears as a viable way to solve NP-hard and non-convex problems like VNF placement. However, these methods suffer from capacity limitation for learning, i.e., single-agent methods \cite{bowling2003multiagent, panait2005cooperative}, and can only handle the environment with only discrete actions (like Deep Q network \textcolor{black}{(DQN)}) or continuous ones (like Deep Deterministic Policy Gradient)\cite{solozabal2019virtual,xiao2019nfvdeep}. Accordingly, some DRL methods are needed in order to handle the network environment with both discrete and continuous actions, while have less capacity limitation for learning to satisfy the network constraint.
	
	\textcolor{black}{\textit{\textbf{Notations}:} 
		Matrices and vectors are denoted by Bold upper-case characters and bold lower-case characters, respectively. $|\mathcal{A}|$ is the cardinality of a set $\mathcal{A}$. $\mathbb{E}[X]$ indicates expected value of $X$.}
	\textcolor{black}{\section{Related Works}} Generally speaking, traditional specific-purpose network architectures cannot provide enough flexibility to adapt to future changes in capacity requirement and very dynamic interconnection of IIoT services and devices. To overcome these problems, NFV can provide efficient resource
	management and flexible network architecture for IoT \cite{shao2018dynamic,farris2018survey}. Using this architecture, communication service providers (CSPs) can provide dynamic and
	adaptive service function chain (SFC) embedding schemes to handle the massive flows of IIoT users in NFV-enabled IIoT networks \cite{fu2019service, 7859379}.
	\subsubsection{VNF Placement and Scheduling}In
	\textcolor{black}{the literature, the authors propose several solutions for VNF placement and the SFC mapping. In \cite{suzuki2020extendable}, the authors employ an efficient coordination algorithm to allocate VNFs into physical networks and route between the VNFs based on reinforcement learning (RL). Although their proposed model adaptively changes the VNF placement based on the network condition, they do not focus on the QoS parameters as well as AoI and the VNF placement cost. In \cite{dieye2018cpvnf}, VNF placement is formulated as an integer linear program and the main goal is to minimize cost of service providers while guaranteeing the QoS requirements. However, they do not consider joint function placement and VNF scheduling in the NFV enabled networks. In \cite{abu2017placement}, a heuristic algorithm is proposed for dynamic placement and scheduling of VNFs and the goal is to minimize the global cost. However, the meta-heuristic search based method cannot optimally handle the stochastic environment with restrict latency requirement since these environments are NP-hard and non-convex and cannot be solved by general Integer Linear Programming (ILP). The VNF Forward Graph Embedding problem is considered in \cite{solozabal2019virtual}, in which the goal is to efficiently map a set of network service VNFs on the physical infrastructure to maximize the remaining resources, while minimizing the power consumption and guaranteeing a specific QoS to each user. They focus on the maximum tolerable latency for services, rather to consider minimizing that. A DRL approach is used in \cite{xiao2019nfvdeep} to jointly minimize the operational cost and maximize the total network throughput for NFV-based networks. However, there are a few works that consider the adaptability of NFV placement and service scheduling in NFV-based IIoT networks.}
	\subsubsection{Age of Information}
	\textcolor{black}{Most previous works deal with minimizing the power consumption, minimizing delay \cite{sun2019age}, minimizing service scheduling cost or maximizing network throughput of VNF placement and service chaining \cite{sun2019age}. However, in most IoT applications such as vehicles, UAVs, health, and sensor monitoring, the real-time updates and fresh data are very important. Therefore, AoI can be a useful metric to measure the freshness of such information \cite{kosta2017age,xu2019optimizing}. In \cite{wu2020uav}, the authors propose a DRL-based algorithm to determine UAVs trajectories during sensing and transmission while guaranteeing the freshness of the sensory data. However, this work is not directly related to VNF placement, hence does not address the issues of VNF placement. In \cite{zhou2019joint,zhou2019minimum}, joint optimal uniform/non-uniform sampling and updating is designed to minimize the average AoI for IoT applications with respect to energy constraint of IIoT devices. Although their proposed model for IIoT network mitigates AoI of the users, considering VNF placement for such a network to decompose network services in order to handle very low latency services would be difficult.} \\
	\textcolor{black}{In the IIoT systems, the large amounts of data generated by smart devices are transmitted over wireless communication networks for analysis and control. To meet the high reliability and real-time demands of IIoT applications, rapid, and frequent data transmission is needed and its performance depends on freshness of status updates \cite{9305697}. In \cite{8972306}, the authors consider an IIoT system, in which randomly generated status updates are sent to the destination through an unreliable channel. Their aim is to minimize the long-term average AoI subject to the average transmission power constraint at the source. In \cite{9273021}, the authors consider a marine IoT system with multiple unmanned surface vehicles (USVs) to monitor the marine environment, while the collected data by sensors on USVs is used to achieve the ubiquitous situation awareness of marine environment. For this purpose, the authors use AoI metric to satisfy freshness of data. The real-time monitoring scenario is also considered in \cite{hsu2020age}, where the collected data is transmitted to the controller and the goal is to minimize the network energy consumption given the constraint on AoI value.} 
	
	\textcolor{black}{Motivated by these considerations, in this paper we propose a new data flow scheduling and VNF placement framework in IIoT networks, in order to minimize the average AoI of the end users as well as transmission and VNF placement cost. Different from the previous works, focusing on the services latency as well as network and radio resource limitation, our proposed model addresses the IIoT challenges more conveniently. More precisely, we formulate the VNF placement, subcarrier and transmit power allocation as a NP-hard and non-convex problem, aiming to minimize the average AoI in IIoT network, then we utilize DRL method to solve our problem. However, the environment of our proposed model consists of both discrete and continuous actions as VNF placement, subcarrier and transmit power allocation. On the other hand, the state-of-the-art DRL methods like DQN and DDPG can only handle discrete and continuous actions, respectively. Therefore, we apply compound action actor critic (CA2C) DRL method which supports both discrete and continuous actions, simultaneously. Finally, we extend our DRL solution to multi-agent method in which multiple agents are cooperating with each other to manage the diverse traffic request in IIoT network with very restrict delay requirement.}
	\subsection{Our Contributions and Works}
	\textcolor{black}{To the best of our knowledge, our work is the first study to  propose a novel AI-based solution to joint radio and NFV cost and AoI minimization problem in IIoT networks by using compound action actor critic algorithm to handle both discrete and continuous optimization variables such as power allocation, subcarrier assignment, and VNF placement.} The main contributions of our paper are summarized as follows:
	\begin{itemize}
		\item We propose a novel framework for the optimal end-to-end data flow scheduling in IIoT applications that jointly
		minimizes  AoI at the destination and total cost of the network.
		\item We formulate a VNF placement and scheduling optimization problem in an IIoT network
		under multiple important constraints such as power budget,
		bandwidth, QoS and AoI metric.
		\item Since our formulated problem is NP-hard and consists of discrete and continuous actions, we employ a single agent (SA) compound-action actor-critic (CA2C) approach which can handle both discrete and continuous actions and achieve near optimal solution for VNF placement and scheduling under the network and user's constraints. 
		\item Furthermore, to overcome the single-agent capacity limitation for learning, we propose a multi-agent DRL method as (MA-CA2C) in which there are multiple agents that are collaborating with each other to satisfy the network constraints. 	
		\item To show the effectiveness of our proposed solution, we conduct simulation using TensorFlow. The simulation results show that the VNF placement and scheduling policy that is obtained by applying the proposed approach provides significant improvement in terms of total network cost and data freshness compared to the
		Greedy approaches.
	\end{itemize}
	
	The remainder of this paper is organized as follows: Section II and  Section III describe system and AoI models used in this paper. Section IV presents problem definition and formulation. Section V presents the baseline DRL algorithms and the proposed MA-CA2C method. Section VI describes the computational complexity and convergence analysis of the proposed solution. The performance evaluation of the proposed scheme is demonstrated in Section VII. Finally, conclusions of this paper are discussed in Section VIII.
	
	\section{System Model}\label{System_Model}
	\subsection{System Model Definition}
	We consider a real-time IIoT system in which a set of IIoT devices collect information from a physical process and send  packets to destination users. A hypervisor allows multiple virtual machines to run on physical layer by virtually sharing their resources, like memory and processing. We consider an SDN controller to manage flow control to improve network management and application performance. $\mathcal{M}$ indicates the set of $M$ IIoT devices. We model a virtual network as a directed graph $\mathcal{G}(\mathcal{N},\mathcal{L})$, where $\mathcal{N}$ is the set of virtual nodes and $\mathcal{L}$ is the set of virtual links, as shown in Fig. \ref{System_Model}. The number of virtual nodes and virtual links in the network are denoted by $|\mathcal{N}|=N$ and $|\mathcal{L}|=L$, respectively. The virtual nodes set consists of source virtual node set, $\mathcal{N}^{\text{S}}$, middle virtual node set, $\mathcal{N}^{\text{R}}$, and destination virtual node set, $\mathcal{N}^{\text{D}}$. There are $M$ IIoT devices that can be connected to every source virtual node and send information to its destination users through middle virtual nodes. There are $\mathcal{J}$ destination users that can be connected to any destination virtual node. The available amount of resources at node $n\in\mathcal{N}$ is indicated by $C_n$ (in Hz) and $B_n$ (in Byte), where $C_n$ and $B_n$ denote the computing and memory capacity at node $n\in\mathcal{N}$, respectively. We define $(\acute{n},n)\in\mathcal{L}$ as a link from virtual node $\acute{n}\in\mathcal{N}^{\text{S}}\cup\mathcal{N}^{\text{R}}$ to virtual node $n\in\mathcal{N}^{\text{R}}\cup\mathcal{N}^{\text{D}}$, where $\acute{n}$ and $n$ are the origin node and the
	destination node of the link $(\acute{n},n)$, respectively. The available bandwidth on each virtual link $(\acute{n},n)\in\mathcal{L}$ is $W_{\acute{n}n}$ (in Hz). 
	\begin{figure*}[h]
	\begin{center}
		\includegraphics[width=6 in]{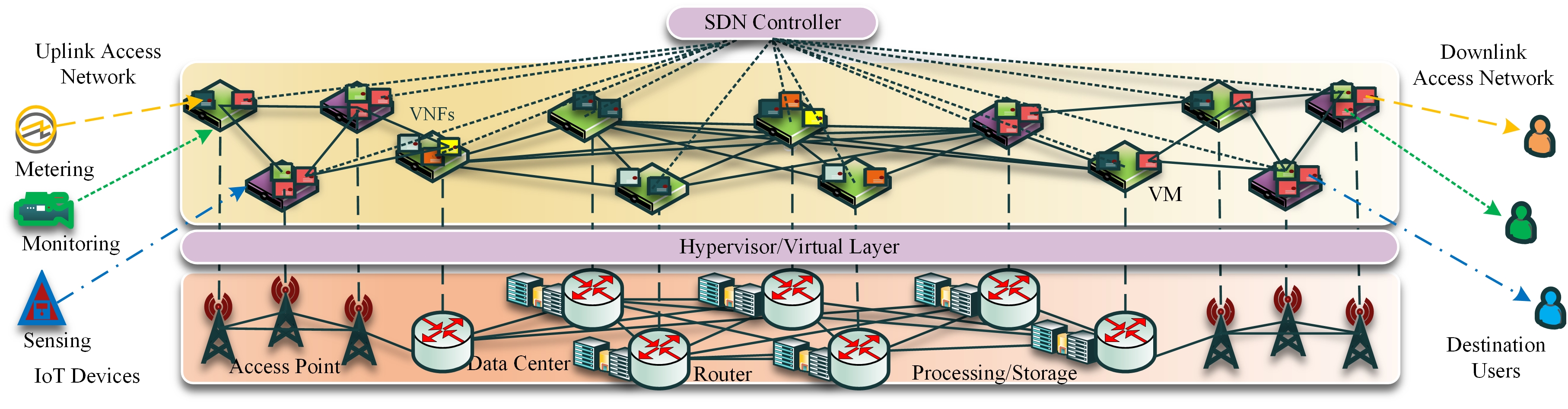} 
		\caption{System model with several candidate VNFs for each service between multiple sources and destinations.}
		\label{System_Model}
	\end{center}\vspace{-.9cm}
\end{figure*}
	Each service provider can provision $\tilde{F}_n$ types of VNFs at virtual node $n$. Let $\mathcal{N}_f$ \textcolor{black}{denote} a set of nodes that can run function $f$. There is a service request set $\mathcal{X}=\{\mathcal{X}_1,\mathcal{X}_2,\dots,\mathcal{X}_k,\dots,\mathcal{X}_K\}$ with $|\mathcal{X}|=K$, in which service $\mathcal{X}_k$ consisting of several VNFs that must be performed sequentially on the packet stream of that service. A SFC for service $k$ is described by the sequence of VNFs as $\mathcal{X}_k=\left( o^1_k,o^2_k,\dots,o^f_k,\dots,o^{F_k}_k\right)$ in which $o^f_k$ is $f^{\text{th}}\in \{1,\dots,f,\dots,F_k\}$ VNF of the chain and $F_k$ is the total number of VNFs for service $k$. The computing, memory, and bandwidth resources required to the $f^{\text{th}}$ VNF of service $k$ are denoted by $c^{fk}_{\acute{n}}$, $b^{fk}_{\acute{n}}$, and $w^{fk}_{\acute{n}n}$, respectively and are defined as, $c^{fk}_{\acute{n}}=\bar{R}_k c_f,~
	b^{fk}_{\acute{n}} =\bar{R}_k b_f,~
	w^{fk}_{\acute{n}n} = \bar{R}_k/\eta_{\acute{n}n}$, where $\bar{R}_k$ denotes the bit rate of service $k$, $c_f$ is the required CPU cycle to execute function $f$ for one bit per second, $b_f$ is the required memory to execute function $f$ for one bit per second, and $\eta_{\acute{n}n}$ indicates the spectral efficiency of link $(\acute{n},n)$. We consider a discrete-time system in which all time slots (TSs) have unit length and is indexed by $t=1,2,\dots$. For service $k$, the updated packets are generated
	by a source, they are then dispersed throughout the network from one node to the next. After entering the packet of service $k$ at node $n$ and performing the required processing on it, the packet is ready to be sent to next node. In each TS, the network must schedule which IIoT service updates its packets at the destination, and which nodes handle the selected services. In other words, in each TS, we allocate the proper VNF to each function of each service and schedule the run time of that function. 
	\begin{definition}\vspace{0.25cm}
		For function $f$ of service $k$ of IIoT device $m$, we define $u^{fkm}_{\acute{n}n}(t)$ as the scheduling action at slot $t$, where $u^{fkm}_{\acute{n}n}(t)=1$ denotes device $m$ is planned to transmit its  packet
		of function $f$ of service $k$ at TS $t$ from node $\acute{n}$ to node $n$ and $u^{fkm}_{\acute{n}n}(t)=0$, otherwise.
	\end{definition}\vspace{0.25cm}
	
	Therefore, to guarantee the QoS,  which is defined as a satisfaction measure of the QoS parameters including the bit rate as well as the AoI, the following constraint should be satisfied for service $k$ of IIoT device $m$:
	\begin{equation}\label{eq-u-1}
		\bar{R}^k_m\leq\sum_{\acute{n}\in\mathcal{N}_f}\sum_{n\in\mathcal{N}_f}u^{fkm}_{\acute{n}n}(t) w^{fk}_{\acute{n}n}(t)\eta_{\acute{n}n}, \forall f,k,m,
	\end{equation}
	where $\bar{R}^k_m$ is the bit rate of service $k$ of IIoT device $m$.
	\subsection{VNF Placement}
	In NFV resource allocation, there are two phases: 1) the VNF placement, 2) the VNF scheduling. In the first phase, the goal is to optimally map SFC requests in NFV-enabled networks. In other words, in the VNF placement, the aim is to choose the optimal locations for a chain of VNFs according to an SFC under limited network resource constraints. We divide all services at TS $t$, $\mathcal{K}(t)$, into three categories: 1) current arrived services, 2) current terminated services and 3) active services. Current arrived service, $\mathcal{K}_R(t)$, are those that arrive in current TS, $\mathcal{K}_R(t)= \left\lbrace k\in\mathcal{K}(t):\psi^{km}(t)=1\right\rbrace$,
	where $\psi^{km}(t)$ is admission control for service $k$ of IIoT device $m$ at TS $t$. We assume that the time is divided into several TSs with equal duration $\delta$ indexed by one integer value, e.g., $t\in \mathbb{N}_+$, where $\mathbb{N}_+$ denotes the set of positive integers. $\mathcal{K}_{T}(t)$ are services that their running time is terminated in TS $t$,  $\mathcal{K}_T(t)=\left\lbrace k\in\mathcal{K}^t:\delta\sum_{f\in\mathcal{F}_k}\sum_{n\in\mathcal{N}_f}\sum_{\acute{t}=t-T^{\text{max}}}^{t}v^{fkm}_n(\acute{t})\geq T^{km} \right\rbrace$, where $T^{km}$ is run-time duration of service $k$ of IIoT device $m$, and $T^{\text{max}}$ indicates the maximum service time among all services which defined as $T^{\text{max}}=\max_{km}\{T^{km}\}$. The binary variable $v^{fkm}_n(t)$ indicates whether function $f$ of service $k$ of IIoT device $m$ is assigned to virtual node $n$ at TS $t$. Active services, $\mathcal{K}_A(t)$ are those that are currently running, $\mathcal{K}_A(t)=\mathcal{K}(t)\setminus\mathcal{K}_R(t)\cup\mathcal{K}_T(t)$. We assume that multiple VNFs of the same type can be run on the same node. However, the bandwidth utilization of the links and the utilization of memory
	and CPU at nodes cannot exceed the available resources, which are ensured as:
	
	\begin{align}\label{eq-u-2}
		&\sum_{t=1}^{\acute{t}}\sum_{m\in\mathcal{M}}\sum_{k\in\mathcal{K}}\sum_{f\in\mathcal{F}_k}w^{fk}_{\acute{n}n}u^{fkm}_{\acute{n}n}(t)\leq r^{\text{BW}}_{\acute{n}n}(\acute{t}),&\\
		&\forall \acute{n}n\in\mathcal{L},\acute{n}\in\mathcal{N}, \acute{t}=1,2,\dots,T,&\nonumber
	\end{align}
	where $r^{\text{BW}}_{\acute{n}n}(t)$ denotes the remaining bandwidth on link $(\acute{n},n)$ at TS $t$  which can
	be updated as:
		\begin{align}\nonumber
		&r^{\text{BW}}_{\acute{n}n}(t+1)=\min\Bigg\lbrace r^{\text{BW}}_{\acute{n}n}(t)-\sum_{m\in\mathcal{M}}\sum_{f\in\mathcal{F}_k}\Big[\\& \sum_{k\in\mathcal{K}_R}w^{fk}_{\acute{n}n}u^{fkm}_{\acute{n}n}(t)-\sum_{k\in\mathcal{K}_T}w^{fk}_{\acute{n}n}u^{fkm}_{\acute{n}n}(t)\Big] ,W_{\acute{n}n}\Bigg\rbrace.
	\end{align}
	
	The allocated memory space on a node cannot exceed the total amount of available memory on that node. As a result:
	\begin{align}\label{eq-2}
		&\sum_{t=1}^{\acute{t}}\sum_{m\in\mathcal{M}}\sum_{k\in\mathcal{K}}\sum_{f\in\mathcal{F}_k}b^{fk}_nv^{fkm}_n(\acute{t})\leq r^{\text{B}}_n(\acute{t}), \forall n\in\mathcal{N},& \\
		&\acute{t}=1,2,\dots,T,&\nonumber
	\end{align}
	where $r^{\text{B}}_n(t)$ represents the remaining memory in node $n$ at TS $t$ and can	be updated as 
	\begin{align}\nonumber
	&r^{\text{B}}_n(t+1)=\min\Big\lbrace r^{\text{B}}_n(t)-\sum_{m\in\mathcal{M}}\sum_{k\in\mathcal{K}_R}\sum_{f\in\mathcal{F}_k}b^{fk}_nv^{fkm}_n(t)\\&+\sum_{k\in\mathcal{K}_T}\sum_{f\in\mathcal{F}_k}b^{fk}_nv^{fkm}_n(t),B_n\Big\rbrace.
\end{align}
	
	The allocated CPU cycles on a node cannot exceed the total amount of available CPU on that node. As a result:
		\begin{equation}\label{eq-3}
	\sum_{t=1}^{\acute{t}}\sum_{m\in\mathcal{M}}\sum_{k\in\mathcal{K}}\sum_{f\in\mathcal{F}_k}c^{fk}_nv^{fkm}_n(\acute{t})\leq r^{\text{C}}_n(\acute{t}), \forall n\in\mathcal{N},
		\end{equation} 
	where $r^{\text{C}}_n(t)$ represents the remaining computing cycle in node $n$ at TS $t$ and can	be updated as 	
	\begin{align}\nonumber
	&r^{\text{C}}_n(t+1)=\min\Big\lbrace r^{\text{C}}_n(t)-\sum_{m\in\mathcal{M}}\sum_{k\in\mathcal{K}_R}\sum_{f\in\mathcal{F}_k}c^{fk}_nv^{fkm}_n(t)+\\&\sum_{m\in\mathcal{M}}\sum_{k\in\mathcal{K}_T}\sum_{f\in\mathcal{F}_k}c^{fk}_nv^{fkm}_n(t),C_n\Big\rbrace.
	\end{align}\vspace{.5cm}
	Function $f$ can only be run by nodes belong to $\mathcal{N}_f$, thus, we have:
	\begin{equation}\label{eq-u-4}
	\sum_{\acute{n}\in\mathcal{N}\setminus \mathcal{N}_f}\sum_{n\in\mathcal{N}}u^{fkm}_{\acute{n}n}(t)=0,~
	\sum_{n\in\mathcal{N}\setminus \mathcal{N}_f}v^{fkm}_n(t)=0.
	\end{equation}\vspace{.5cm}
	The VNF nodes can not fragment their flows into multiple links. The following condition must be met to ensure that only one link is assigned to the output of each function:
	\begin{equation}\label{eq-u-5}
	\sum_{\acute{n}\in\mathcal{N}_f}\sum_{n\in\mathcal{N}}u^{fkm}_{\acute{n}n}(t)=1.
	\end{equation}\vspace{.5cm}
	To ensure that for each function of a service, either none or exactly one instance of VNF is assigned to a node, the following constraint must be satisfied:
	\begin{equation}\label{eq-u-55}
	\sum_{n\in\mathcal{N}_f}v^{fkm}_n(t)\leq 1,\forall f,k,m,t.
	\end{equation}
	\subsection{Routing}
	Each service $k\in \mathcal{K}$ is characterized by a source IIoT device $m$ that wants to transmit a flow of data toward a destination user $j$. To impose a single path flow balance, the following flow conservation constraint must be satisfied:
	\begin{align}\label{eq-u-6}
	&\sum_{f\in\mathcal{F}_k}\sum_{n\in\mathcal{N}}u^{fkm}_{\acute{n}n}(t)-\sum_{f\in\mathcal{F}_k}\sum_{n\in\mathcal{N}}u^{fkm}_{n\acute{n}}(t)=\\\nonumber&\begin{cases}
		1, & \text{if $\acute{n}\in\mathcal{N}^{\text{S}}$},\\
		-1, & \text{if $\acute{n}\in\mathcal{N}^{\text{D}}$},\\
		0, & \text{otherwise}.
	\end{cases} 
	\hspace{0.5cm} \forall k,m,t,\acute{n}.
	\end{align}
	
	To guarantee that if function $f$ of service $k$ of IIoT device $m$ is placed at node $\acute{n}$ then the traffic flow will be routed through that node, the following constraints must be always satisfied:
	\begin{equation}\label{eq-u-7}
		\sum_{n\in\mathcal{N}}u^{fkm}_{\acute{n}n}(t)\geq v^{fkm}_{\acute{n}}(t), \forall f,k,\acute{n},t.
	\end{equation}
	\subsection{The VNF Scheduling}
	The VNF scheduling is done in the second phase and is based on the first phase results, i.e, VNF placement. Therefore, a different placement scheme will lead to a different solution for VNF scheduling. Based on a given placement, VNF scheduling aims to find the best scheduling scheme to execute VNFs at the corresponding virtual nodes. We assume that $\tau^{(f+1)km}_n$ is starting time of VNF $(f+1)$ of the $k^{\text{th}}$ service for IIoT device $m$ which is allocated to node $n$. VNF $(f+1)$ of service $k$ in node $n$ can start its service if and only if execution of its preceding VNF $f$ is finished. Let  $\tilde{\tau}^{fkm}_{\acute{n}n}$ denote the propagation delay between virtual nodes $\acute{n}$ and $n$ and it can be calculated as, $\tilde{\tau}^{fkm}_{\acute{n}n}=d_{\acute{n}n}/c$, where $c$ is light speed and $d_{\acute{n}n}$ is distance between nodes $\acute{n}$ and $n$. Since the virtualized nodes are mapped to the physical ones, where they can be far from each other, the propagation delay can be notable \cite{7859379,7842225}. The time it takes to upload data through the link (transmission delays) and the processing time of each VNF denoted by $\bar{\tau}^{fkm}_{\acute{n}n}(t)$ and $\hat{\tau}^{fkm}_{\acute{n}}(t)$, respectively. They can be calculated as $\bar{\tau}^{fkm}_{\acute{n}n}(t)=\bar{R}_k/ w^{fk}_{\acute{n}n}\eta_{\acute{n}n}$ and
	$\hat{\tau}^{fkm}_{\acute{n}}(t)=\bar{R}_kc^{fk}_{\acute{n}}/C_{\acute{n}}$. Therefore, the run-time of VNF $f$ of service $k$ for IoT device $m$ at virtual node $n$ can be determined as: 
	\begin{align}\label{eq-u-8}
	&\tau^{(f+1)km}_n(t+1)=\sum_{\substack{\acute{n}\in\mathcal{N},\acute{n}\neq n}}v^{fkm}_{\acute{n}}(t)\tau^{fkm}_{\acute{n}}(t)+\\\nonumber&\sum_{\substack{\acute{n}\in\mathcal{N},\acute{n}\neq n}}v^{fkm}_{\acute{n}}(t)\hat{\tau}^{fkm}_{\acute{n}}(t)+\sum_{\substack{\acute{n}\in\mathcal{N},\acute{n}\neq n}}u^{fkm}_{\acute{n}n}(t)\tilde{\tau}^{fkm}_{\acute{n}n}+\\\nonumber&\sum_{\substack{\acute{n}\in\mathcal{N},\acute{n}\neq n}}u^{fkm}_{\acute{n}n}(t)\bar{\tau}^{fkm}_{\acute{n}n}(t),
	\end{align}
	
	To guarantee that VNF $f$ of service $k$ of device $m$ completes its process in the allocated TS, the summation of the propagation delay $\tilde{\tau}^{fkm}_{\acute{n}n}$, the processing time $\hat{\tau}^{fkm}_{\acute{n}}(t)$, and transmission delay, $\bar{\tau}^{fkm}_{\acute{n}n}(t)$ with $\tau^{fkm}_n(t)$ should not exceed duration time of one slot, $\delta$:
	\begin{align}\label{eq-u-77}
	\tau^{fkm}_n(t)+\hat{\tau}^{fkm}_n(t)+\bar{\tau}^{fkm}_{\acute{n}n}(t)+\tilde{\tau}^{fkm}_{\acute{n}n}(t) \leq \delta,
	\end{align}
	
\begin{figure*}[h]
	\begin{center}
		\includegraphics[width=6 in]{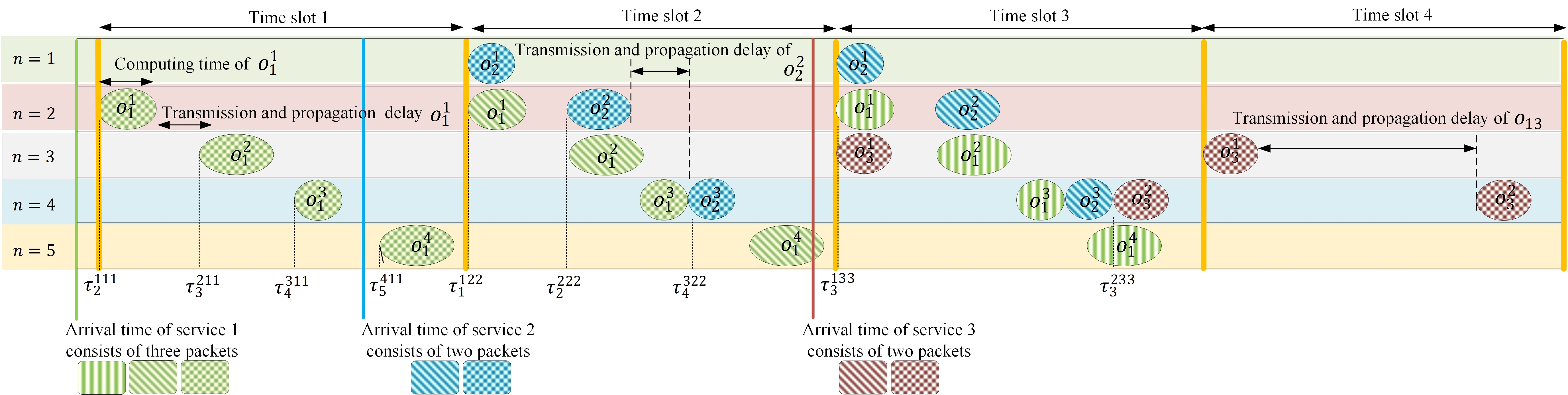} 
		\caption{Scheduling of services 1, 2 and 3: At beginning of each TS, the scheduler assigns optimal node to each function in order to guarantee the QoS of all services.} 
		\label{Service3}
	\end{center}\vspace{-.9cm}
	\end{figure*}

	An example for three services, indexed by $k=\{1,2,3\}$ correspond to IIoT devices indexed by $m=\{1,2,3\}$, is illustrated in Fig. \ref{Service3}. First service arrives at the time $\tilde{t}_1$ and requests VNFs $\{o^1_{1},o^2_{1},o^3_{1},o^4_{1}\}$ as shown in Fig. \ref{Service3}. As can be seen from Fig. \ref{Service3}, the VNFs $o^1_{1},o^2_{1},o^3_{1}$, and $o^4_{1}$ are assigned to nodes $2,3,4$, and $5$, respectively. 
	Similarly, second service arrives at a time $\tilde{t}_2$ with VNFs $\{o^1_{2},o^2_{2},o^3_{2}\}$ which were assigned to nodes $1,2$, and $4$, respectively. Finally, third service arrives at a time $\tilde{t}_3$ with VNFs $\{o^1_{3},o^2_{3}\}$ which were assigned to nodes $3$, and $4$, respectively.
	\subsection{Access Networks}
	\subsubsection{Uplink Access Network}
	We assume that the total bandwidth of uplink (UL) access network is divided into $H$ subcarriers where $\mathcal{H}=\{1,\dots,h,\dots,H\}$ indicates the index set of all subcarriers \cite{cheung2013achieving,abedi2016limited}. The total allocated bandwidth does not exceed a given total bandwidth budget, e.g. $\sum_{m\in\mathcal{M}}\sum_{n\in\mathcal{N}}\sum_{h\in\mathcal{H}}\sum_{k\in\mathcal{K}}
	\rho^{hk}_{mn}\bar{w}_{h}\leq\bar{W}$ where $\bar{w}_{h}$ denotes the bandwidth of subcarrier $h$, $\bar{W}$ is total bandwidth budget at UL access network. $\rho^{hk}_{mn}$ denotes an integer variable for assigning subcarriers to service $k$ and link between IIoT device $m$ and node $n$, where $\rho^{hk}_{mn}=1$ if subcarrier $h$ is allocated to service $k$ and link between IIoT device $m$ and node $n$, and otherwise $\rho^{hk}_{mn}=0$. To guarantee the end-to-end QoS, the following constraints should be satisfied for bit rate of service $k$ of IIoT device $m$, $R^k_{m}$ in UL access network, as well:
	\begin{equation}\label{eq-R}
	\bar{R}^k_{m}\leq R^k_{m}=\sum_{m\in\mathcal{M}}\sum_{n\in\mathcal{N}}\sum_{h\in\mathcal{H}}\rho^{hk}_{mn}\bar{w}_{h}\log_2( 1+\frac{p^{hk}_{mn}g^{h}_{mn}}{\sigma^2}),
	\end{equation}
	where $p^{hk}_{mn}$ and $g^{h}_{mn}$ denote the power allocated to the service $k$ of IIoT device $m$ on subcarrier $h$ and channel gain between IIoT device $m$ and node $n$ on subcarrier $h$, respectively. $\sigma^2$ denotes the variance of additive white Gaussian noise. Each subcarrier can only be assigned to at most one IIoT device:
	\begin{equation}\label{eq-rho}
		\sum_{m\in\mathcal{M}}\sum_{n\in\mathcal{N}}\sum_{k\in\mathcal{K}}\rho^{hk}_{mn}\leq 1, \forall h.
	\end{equation}
	
	The access network delay consists of transmission and propagation delays, which are calculated as $\bar{\pi}^{k}_{mn}=\bar{R}^k_m/ \sum_{h\in\mathcal{H}}\rho^{hk}_{mn}\bar{w}_{h}R^k_{m},
	\hat{\pi}_{mn}=d_{mn}/c$. The access delay should not exceed the threshold value, $\epsilon^k_m$
	\begin{align}\label{eq-access-delay-1}
		\hat{\pi}_{mn}+\bar{\pi}^{k}_{mn}\leq \epsilon^k_m.
	\end{align}
	\subsubsection{Downlink Access Network}
	We assume that the total bandwidth of downlink (DL) access network is divided into $\check{H}$ subcarriers where $\check{\mathcal{H}}=\{1,\dots,h,\dots,\check{H}\}$ denotes the index set of all subcarriers. The total allocated bandwidth do not exceed a given total bandwidth budget, e.g. $\sum_{j\in\mathcal{J}}\sum_{n\in\mathcal{N}}\sum_{\check{h}\in\mathcal{\check{H}}}\sum_{k\in\mathcal{K}}\rho^{\check{h}k}_{nj}\textcolor{black}{\check{w}_{\check{h}}}\leq\check{W}$ where $\check{w}_{\check{h}}$ \textcolor{black}{denotes} the bandwidth of subcarrier $\check{h}$, $\check{W}$ is total bandwidth budget at access network and $\rho^{\check{h}k}_{nj}$ \textcolor{black}{denotes} the integer variable for assigning subcarriers to service $k$ and link between destination node $n$ and destination user $j$, where $\rho^{\check{h}k}_{nj}=1$ if subcarrier $\check{h}$ is allocated to service $k$ and link between destination node $n$ and destination user $j$, and otherwise $\rho^{\check{h}k}_{nj}=0$. To guarantee the end-to-end QoS, the following constraints should be satisfied for bit rate of service $k$ of destination user $j$, $\check{R}^k_j$ in access network, as well:
	
	\begin{equation}\label{eq-R-dl}
	\bar{R}^k_{j}\leq \check{R}^k_j=\sum_{j\in\mathcal{J}}\sum_{n\in\mathcal{N}}\sum_{\check{h}\in\mathcal{\check{H}}}\rho^{\check{h}k}_{nj}\check{w}_{\check{h}}\log_2\left( 1+\frac{p^{\check{h}k}_{nj}g^{\check{h}}_{nj}}{\sigma^2}\right),
	\end{equation}
	where $\bar{R}^k_{j}$ is the bit rate of service $k$ of user $j$, $p^{\check{h}k}_{nj}$ and $g^{\check{h}}_{nj}$ denote the power allocated to the service $k$ of destination user $j$ on subcarrier $\check{h}$ and channel gain between destination node $n$ and destination user $j$ on subcarrier $\check{h}$, respectively. Each subcarrier can only be assigned to at most one IIoT device:	
	\begin{equation}\label{eq-rho-dl}
	\sum_{j\in\mathcal{J}}\sum_{n\in\mathcal{N}}\sum_{k\in\mathcal{K}}\rho^{\check{h}k}_{nj}\leq 1, \forall \check{h},
	\end{equation}
	
	The access network delay consists of transmission and propagation delays, which are calculated similar to (17).\vspace{-0.1cm}
	\section{AoI Model for the Proposed System}
	In this section, the AoI is adopted to formulate the freshness of the sensing data collected by the IIoT devices. 
	\begin{definition}\vspace{.25cm}
		For service $k$ of IIoT device $m$, let $X_n^{km}(t)$ and $Y_n^{km}(t)$ be the AoI of the  packet of service $k$ of IIoT device $m$ at source node $n$ and destination node $n$ at TS $t$, respectively. We define $\Delta_j^{km}(t)$ as the AoI of the  packet of service $k$ of IIoT device $m$ at the
		destination user $j$ at TS $t$. 
		Let $\hat{X}_n^{km}$ and $\hat{\Delta}_j^{km}$ denote the upper limits of the AoI of the  packet of service $k$ of IIoT device $m$ at node $n$ and the destination user $j$, respectively. Considering the real-time application of the proposed scheme, it is clear that $\hat{X}_n^{km}$ and $\hat{\Delta}_j^{km}$ are finite.
	\end{definition}\vspace{.25cm}
	Note that $X_n^{km}(t)$, $Y_n^{km}(t)$, and $\Delta_j^{km}(t)$ depend on how the scheduling is done. We assume that $X_n^{km}(t)$ is equal to $0$ at $t=0$. For service $k$ of IIoT device $m$, if there is a  packet arriving at source node $n$ at TS $t$, $X_n^{km}(t)$ \textcolor{black}{will be equal to sum of propagation and transmission delay from node $m$ to node $n$, i.e} $\hat{\pi}_{mn}+\bar{\pi}^{k}_{mn}$, otherwise, it will
	increase by $\delta$ (due to the one slot transmission). Therefore, the AoI of the  packet of service $k$ of device $m$ at source node $n$, $X_n^{km}(t+1)$, can be updated as follows:	
	\begin{align}
	&X_n^{km}(t+1)=\\\nonumber&
	\begin{cases}
		\hat{\pi}_{mn}+\bar{\pi}^{k}_{mn},&\text{If}\rho^{hk}_{mn}(t)=1,\\
		\min\{X_n^{km}(t)+\delta,\hat{X}_n^{km}\},&\text{otherwise},
	\end{cases}
	\end{align}
	where $\rho^{hk}_{mn}(t)=1$ means in the previous slot $t$, subcarrier $h$ was allocated to service $k$ of IIoT device $m$ to deliver the updated packet to source node $n$. Similarly, for service $k$ of IIoT device $m$, if there is a  packet arriving at destination node $n$ at TS $t$, $Y_n^{km}(t)$ will be $\hat{\pi}_{mn}+\bar{\pi}^{k}_{mn}+\delta$, \textcolor{black}{since in the previous TS $t-1$, a successful transmission from IIoT device $m$ to source node $n$ has happened and the term $\delta$ is also added because the service routing has happened in TS $t$. If in the TS $t-1$ no packet is transmitted to source node $n$ from IIoT device $m$ and a successful VNF placement occurs,  then the AoI of destination node $n$ will be equal to one TS plus $X_n^{km}(t)$,} otherwise, it will increase by $\delta$. Therefore, the AoI of the  packet of service $k$ of device $m$ at destination node $n$, $Y_n^{km}(t+1)$, can be updated as follows:
	\begin{align}
	&Y_n^{km}(t+1)=
	\begin{cases}
		\hat{\pi}_{mn}+\bar{\pi}^{k}_{mn}+\delta,&\text{If}~\Theta_1,\\
		\min\{X_n^{km}(t)+\delta,\hat{X}_n^{km}\},&\text{If}~\Theta_2,\\
		\min\{Y_n^{km}(t)+\delta,\hat{Y}^n_{km}\},&\text{otherwise},
	\end{cases}
	\end{align}
	where $\Theta_1=\rho^{hk}_{mn}(t-1)\varrho_{km}(t)=(F_k)^2$, $\Theta_2=\rho^{hk}_{mn}(t-1)=0$. $\rho^{hk}_{mn}(t-1)=1$ means in the two previous slots $t-1$, subcarrier $h$ was allocated to service $k$ of IIoT device $m$ to deliver its updated packet to source node $n$ and $\varrho_{km}(t)=\left( \sum_{f\in\mathcal{F}_k}\sum_{\acute{n}\in\mathcal{N}\setminus \mathcal{N}_f}\sum_{n\in\mathcal{N}}u^{fkm}_{\acute{n}n}(t)\right)\\\left( \sum_{f\in\mathcal{F}_k}\sum_{n\in\mathcal{N}}v^{fkm}_n(t)\right)$ where $\varrho_{km}(t)=(F_k)^2$ means in the previous slot $t$, $F_k$ nodes and $F_k$ links were allocated to $F_k$ functions of service $k$ of IIoT device $m$ to deliver the update packet to destination node $n$. Finally, at destination user $j$, for the AoI of service $k$ of device $m$, if no destination node delivers a fresh packet to destination user $j$ at TS $t$, then the AoI of service $k$ of device $m$, $\Delta_j^{km}(t+1)$ will increase by $\delta$, otherwise, the  information of service $k$ of device $m$ is updated according to how we receive a packet from destination node $n$, as depicted in (\ref{eq_21}).\vspace{-0.2cm}
	\begin{strip}
		\begin{align}\label{eq_21}
		\Delta_j^{km}(t+1)=\begin{cases}
			\hat{\pi}_{mn}+\bar{\pi}^{k}_{mn}+\delta+\check{\pi}^{k}_{nj}+\tilde{\pi}_{nj},&\text{If}~\rho^{hk}_{mn}(t-2)\varrho_{km}(t-1)\rho^{\check{h}k}_{nj}(t)=(F_k)^2,\\
			\min\{X_n^{km}(t)+\delta+\check{\pi}^{k}_{nj}+\tilde{\pi}_{nj},\hat{\Delta}_{km}\},&\text{If}~\rho^{hk}_{mn}(t-2)=0 \&\varrho_{km}(t-1)\rho^{\check{h}k}_{nj}(t)=(F_k)^2,\\
			\min\{Y_n^{km}(t)+\check{\pi}^{k}_{nj}+\tilde{\pi}_{nj},\hat{\Delta}_{km}\},&\text{If}\rho^{hk}_{mn}(t-2)=0 \&\varrho_{km}(t-1)=0\&\rho^{\check{h}k}_{nj}(t)=1,\\
			\min\{\Delta_{km}(t)+\delta,\hat{\Delta}_{km}\},&\text{otherwise},
		\end{cases}
		\end{align}
	\end{strip}
	\textcolor{black}{where $\Delta_{km}(t)=t-U(t)$, $U(t)$ denotes the timestamp of the most recently received status packet at time $t$}, $\rho^{hk}_{mn}(t-2)=1$ means in the three previous slots $t-2$ a subcarrier was allocated to service $k$ of IIoT device $m$ at UL network to deliver the update packet to source node $n$, $\varrho_{km}(t-1)=(F_k)^2$ means in the two previous slots $t$, $F_k$ nodes and $F_k$ links were allocated to $F_k$ functions of service $k$ of IIoT device $m$ to deliver the update packet to destination node $n$, and $\rho^{\check{h}k}_{nj}(t)=1$ means in the previous slot $t$, subcarrier $\check{h}$ was allocated to service $k$ of IIoT device $m$ at DL network to deliver the update packet to destination $j$.
	\section{Problem Definition and Formulation}
	We aim to manage VNFs of all services of IoT devices, so as to jointly minimize the VNF placement, scheduling cost and the maximum of AoIs of devices.
	\begin{definition}\vspace{.25cm}
		Let $\beta^{fkm}_{\acute{n}n}(t)$ \textcolor{black}{denote} the cost for forwarding the  information of function $f$ of service $k$ of device $m$ over virtual link $(\acute{n},n)$ and $\alpha^{fkm}_n(t)$ the cost for processing the information of function $f$ of service $k$ of device $m$ at node $n$. We also define $\zeta^{hk}_{mn}(t)$, and $\zeta^{\check{h}k}_{nj}(t)$ as the cost for forwarding the  information of service $k$ of device $m$ to node $n$ over wireless link $h$ at TS $t$, the cost for forwarding the  information of service $k$ of device $n$ to user $j$ over wireless link $\check{h}$ at TS $t$, respectively.
	\end{definition}\vspace{.25cm}
	We formulate the optimization problem as the minimization of the sum of \textcolor{black}{VNF} placement and scheduling cost in addition to the average of AoIs of devices in respect to total transmit power at each node and device, each link bandwidth and CPU and memory capacity of each node constraints: 
	
	\begin{align}\label{Opt}
	&\min_{\boldsymbol{u},\boldsymbol{v},\boldsymbol{\rho}} \xi\bar{\Delta}+\xi_1\Lambda_1+\xi_2\Lambda_2+\xi_3\Lambda_3+\xi_4\Lambda_4\\&\nonumber
	\bar{\Delta}=\sum_{k,m}{\bar{\Delta}_{km}}\\&\nonumber
	\bar{\Delta}_{km}=\lim_{\acute{\tau}\to\infty}\frac{1}{\acute{\tau}}\int_{0}^{\acute{\tau}}{\Delta}_{km}(t)dt\nonumber\\&
	\Lambda_1=\sum_{t}\sum_{k\in\mathcal{K}}\sum_{f\in\mathcal{F}_k}\sum_{m\in\mathcal{M}}\sum_{(\acute{n},n)\in\mathcal{L}}u^{fkm}_{\acute{n}n}(t)\beta^{fkm}_{\acute{n}n}(t)\nonumber\\&
	\Lambda_2=\sum_{t}\sum_{k\in\mathcal{K}}\sum_{f\in\mathcal{F}_k}\sum_{m\in\mathcal{M}}\sum_{(\acute{n},n)\in\mathcal{L}}v^{fkm}_n(t)\alpha^{fkm}_n(t)\nonumber\\&
	\Lambda_3=\sum_{t}\sum_{k\in\mathcal{K}}\sum_{h\in\mathcal{H}}\sum_{m\in\mathcal{M}}\sum_{n\in\mathcal{N}}\zeta^{hk}_{mn}(t)\rho^{hk}_{mn}(t)(\bar{\omega}_h+p^{hk}_{mn})\nonumber\\&
	\Lambda_4=\sum_{t}\sum_{k\in\mathcal{K}}\sum_{\check{h}\in\mathcal{\check{H}}}\sum_{j\in\mathcal{J}}\sum_{n\in\mathcal{N}}\zeta^{\check{h}k}_{mn}(t)\rho^{\check{h}k}_{mn}(t)(\check{\omega}_{\check{h}}+p^{\check{h}k}_{mn})\nonumber\\&
	s.t.(\ref{eq-u-1}),(\ref{eq-u-2}),(\ref{eq-2}),(\ref{eq-3}),(\ref{eq-u-4})\nonumber
	\end{align}
	where $\bar{\Delta}$ is the average AoI of destination users, and $\bar{\Delta}_{km}$ is the average AoI of service $k$ of IIoT device $m$, $\Lambda_1$ indicates cost for executing all functions of services,
	$\Lambda_2$ denotes cost for forwarding all functions of services, $\Lambda_3$ indicates the cost for forwarding the  information of service $k$ of device $m$ over link $h$ and $\Lambda_4$ denotes the cost for forwarding the  information of service $k$ of device $m$ over link $\check{h}$,   $\xi,\xi_1,\dots,\xi_4$ are the weights which act as balancing parameters among various parts of our objective function.\\
	Since the remaining resources and the channel conditions are dynamically varying (i.e., stochastic environment), (\ref{Opt}) is a dynamic optimization
	problem which can be formulated as the Markov decision process (MDP). Moreover, due to the non-convexity of some constraints and the discrete sets in the constraints, this problem is a non-convex and NP-hard optimization problem. Traditional methods such as static
	optimization and game theory can not obtain the optimal solutions, especially in large scale networks. In addition, heuristic algorithms can not necessarily reach close-to-optimal results due to lack of strict theoretical proof. To overcome these drawbacks, the DRL-based algorithms can be utilized to obtain the near-optimal solution for such problems. 
	
	\section{DRL-based Solution}
	\textcolor{black}{Except in a few cases in wireless resource allocation, where the optimization problems can be formulated as a convex optimization problem, most optimization problems are non-convex and NP-hard because they include joint power allocation and spectrum assignment. Therefore, the traditional algorithms such as convex optimization cannot optimally solve these problems with polynomial time complexity, especially in large scale networks. In the case of heuristic algorithms, it is difficult to reach close-to-optimal results due to lack of strict theoretical proof. To tackle these challenges, DRL-based algorithms can be effectively used to solve such problems. Many recent works show that the machine-learning-based resource allocation outperforms conventional methods \cite{8943940,sun2019age,8103164}.} In this section, we first introduce some basics about DRL and then introduce the proposed MA-CA2C algorithm. In the traditional RL, agents interact with the environment and learn to take the action that would yield the most cumulative reward. A typical RL problem is modeled as an MDP. An MDP mainly consists of a tuple as $\big \{s(t), a(t), r(t), s(t+1)\big\}$. At each decision epoch (TS) $t$, agent is at state $s(t)$ and takes an action $a(t)$ based on a policy and causes the environment transitions to new state $s(t+1)$ while it receives an immediate reward $r(t)$. Basically, the aim of MDP is to train an agent to find a policy \textcolor{black}{$\pi$ which maps the state $s(t)$ to action $a(t)$, i.e., $a(t)=\pi(s(t))$, and} returns the maximum expected cumulative discounted future reward as $r(t)=\mathbb{E}\{\sum_{l=t}^{T}\gamma(l-t)r(t+l)\}$ by taking a series of actions in one or more states, where $\gamma\in[0,1]$ is a discount factor \textcolor{black}{indicating how future rewards is important} and $T$ is the number of decision TSs. In traditional Q-learning, the Q-table is used to store \textcolor{black}{state-action values pairs}. In large scale systems, like the considered system model in this paper, the state and action spaces are very large and it leads to a large $Q$-table, therefore, using one table to store all action values is impractical.
	\subsection{Baseline Algorithms}
	\subsubsection{Deep Q-Network}
	Recently, DRL is introduced as a promising technique to overcome the mentioned challenges in RL. In this algorithm, deep neural networks (DNNs) are used as the function approximators to predict Q-values. Therefore, DRL not only enhances the performance of traditional RL schemes, but also can manage continuous state and action spaces. The $Q$-function estimated by the neural network is represented by $Q(s(t),a(t);\boldsymbol{\theta}(t))$ where the parameter $\boldsymbol{\theta}(t)$ has the weights for the neural network and its updated value is used to train the neural network and approximate the real values of $Q$ \cite{wu2020uav,fu2019service}. By minimizing the loss function $L(\boldsymbol{\theta}(t))$ at each iteration, the deep $Q$-function is trained to learn the best fitting, where the loss function is \textcolor{black}{the expectation of mean squared error of estimated value from the target value and }can be expressed as $L(\boldsymbol{\theta}(t))=\mathbb{E}[(y(t)-Q(s(t),a(t);\boldsymbol{\theta}(t)))^2]$ \cite{wu2020uav}, where $y(t)=R(t)+\gamma\max_{a(t+1)}Q(s(t+1),a(t+1);\boldsymbol{\theta}(t))$ is the target value and $a(t+1)$ indicates the action generated by the DNN in TS $t+1$, given the state $s(t+1)$.	Although, the DNN provides accurate approximation, it may lead to the divergence of the learning algorithm or very ineffective learning because of the non-stationary target values and samples correlation. To overcome the non-stationary and samples correlation challenges, a pair of techniques, namely Fixed Target Network \cite{mnih2015human} and Experience Replay \cite{hou2017novel} are utilized, respectively. By adopting these techniques, the loss function can be written as
	\begin{align}\label{00-112}\nonumber
		&L(\boldsymbol{\theta}(t))=\mathbb{E}_{D}[(R(t)+\gamma\max_{a(t+1)}Q(s(t+1),a(t+1);\acute{\boldsymbol{\theta}}(t))\\&-Q(s(t),a(t);\boldsymbol{\theta}(t)))^2],
	\end{align}
	where $\acute{\boldsymbol{\theta}}(t)$ denotes the parameters of the target network, and the expectation is taken over the mini-batches sampled from $D$.
	The DQN with target network and experience replay can only handle the environment with discrete action spaces \cite{mnih2015human}. \textcolor{black}{Since} the proposed scheme in this paper consists of continuous and discrete actions as power allocation and VNF placement, respectively, \textcolor{black}{the DQN method can not optimally solve our proposed problem.}
	\subsubsection{Deep Deterministic Policy gradient Deep RL Algorithm}
	Model-free deep RL approaches like DQN, dueling DQN and double DQN are value-based algorithms in which the Q-values are estimated with lower variance. These solutions can not handle the problems with continuous action spaces. To overcome this challenge, policy gradient-based RL approaches can be used, which can deal with problems with continuous action space by learning deterministic/stochastic policies. In these methods, the aim is to optimize a policy based on the gradient of the expected reward. However, these methods have very slow convergence \cite{lillicrap2015continuous}. To tackle this problem, the Deep Deterministic Policy Gradient abbreviated as DDPG is utilized to integrate both properties of policy-based and value-based algorithms in order to deal with continuous and large state/action spaces\cite{yang2020dynamic, lillicrap2015continuous}. In this algorithm, there are two separated neural networks, an actor and a critic network with parameters $\boldsymbol{\omega}$ and $\boldsymbol{\theta}$, respectively. The actor outputs continuous actions via a deterministic policy $\pi(a(t)|s(t);\boldsymbol{\omega}(t))$, and the critic evaluates the action taken by the actor via a $Q(s(t),a(t);\boldsymbol{\theta}(t))$. Since obtaining the explicit expressions of $\pi(a(t)|s(t);\boldsymbol{\omega}(t))$ and $Q(s(t),a(t);\boldsymbol{\theta}(t))$ is difficult, the actor approximates a deterministic policy $\pi(a(t)|s(t);\boldsymbol{\omega}(t))$ and the critic approximates value function $Q(s(t),a(t);\boldsymbol{\theta}(t))$. For the current state $s(t)$, the actor generates an action $a(t)$ based on a deterministic policy $\pi(a(t)|s(t);\boldsymbol{\omega}(t))$. Then, critic calculates the loss function based on the estimated Q-value with $Q(s(t),a(t);\boldsymbol{\theta}(t))=\mathbb{E}_{a(t)\sim\pi(a(t)|s(t);\boldsymbol{\omega}(t))}[R(t)|s(t),a(t)]$. \textcolor{black}{The expectation is taken over the action $a(t)$ based on the deterministic policy $\pi$, i.e., $a(t)\sim\pi(a(t)|s(t);\omega(t))$}. The loss function of the actor is defined as $-J(\boldsymbol{\omega}(t))=-\mathbb{E}[Q(s(t),a(t);\boldsymbol{\theta}(t))]$. The actor is updated by applying
	the policy gradient method as $\nabla_{\boldsymbol{\omega}}J(\boldsymbol{\omega}(t))=\mathbb{E}[\nabla_{\boldsymbol{\omega}}\log\pi(a(t)|s(t);\boldsymbol{\omega}(t))Q(s(t),a(t);\boldsymbol{\theta})]$. Then, the gradient descent is used to update $\boldsymbol{\omega}(t)$:
	\begin{align}
		\boldsymbol{\omega}(t+1)=\boldsymbol{\omega}(t)-\alpha_{a}(-\nabla_{\boldsymbol{\omega}}J(\boldsymbol{\omega}(t))),
	\end{align}
	where $\alpha_a$ indicates the learning rate of the actor. For the continuously differentiable loss function with respect to $\boldsymbol{\theta}$, the gradients of the loss function can be used for updating the parameters $\boldsymbol{\theta}$ of the critic DNN as follows:
	\begin{align}\nonumber
	&\Delta\boldsymbol{\theta}=\alpha_c[R(t)+\gamma\max_{a(t+1)}Q(s(t+1),a(t+1);\acute{\boldsymbol{\theta}})-\\&Q(s(t),a(t);\boldsymbol{\theta})]\nabla_{\boldsymbol{\theta}}Q(s(t),a(t);\boldsymbol{\theta}),
	\end{align}
	where $\alpha_c$ is critic learning rate. Moreover, since a mini-batch of size $I$ is used to update the parameter $\boldsymbol{\theta}$, we have
		\begin{align}\nonumber
	&\Delta\boldsymbol{\theta}=\frac{\alpha_c}{I}\sum_{i=1}^{I}[R^i(t)+\gamma\max_{a^i(t+1)}Q(s^i(t+1),a^i(t+1);\acute{\boldsymbol{\theta}})\\&-Q(s^i(t),a^i(t);\boldsymbol{\theta})]\nabla_{\boldsymbol{\theta}}Q(s^i(t),a^i(t);\boldsymbol{\theta}),
	\end{align}
	where index $i$ refers to sample $i$. 
	\subsubsection{Compound-Action Actor Critic DRL Algorithm}
	DRL algorithms are suitable for finding the optimal policies in  problems with high-dimensional state spaces and purely discrete or purely continuous action spaces. Therefore, these algorithms cannot be applied directly to the problems with actions involving both continuous and discrete variables. Since our optimization problem includes both continuous and discrete optimization variables, we can not directly use these algorithms. To overcome this challenge, we utilize compound algorithm proposed by \cite{hu20201} which consists of both DDPG and DQN algorithm for handling continuous and discrete actions, respectively. \textcolor{black}{The environment of our proposed model consists of both discrete and continuous actions as VNF placement, subcarrier and transmit power allocation, while the state-of-the-art DRL methods like DQN and DDPG can only handle discrete and continuous actions, respectively. To overcome this challenge, we apply CA2C which supports both discrete and continuous actions simultaneously.} Considering our optimization problem, we define the state space $\mathcal{S}$, the action space $\mathcal{A}$, and the reward function $R$ as follows:
	\begin{itemize}
		\item \textbf{State:} We define the state space as a vector of residual resources in VNF nodes, residual bandwidth on the links, the UL/DL channel conditions, the number of requested services and the desired bit rate of requested services. As a result, the state $s(t)\in\mathcal{S}$ at decision TS $t$ can be written as	
		$s(t)=\left\lbrace \textbf{r}^{BW}(t), \textbf{r}^{M}(t), \textbf{r}^{C}(t),\textbf{g}(t),\bar{\textbf{R}}(t)\right\rbrace$ where $\textbf{r}^{BW}(t)=[r^{BW}_{\acute{n}n}(t)],\textbf{r}^{M}(t)=[r^{M}_{n}(t)],\textbf{r}^{C}(t)=[r^{C}_{n}(t)],\textbf{g}(t)=[g^{h}_{mn}(t)],\bar{\textbf{R}}(t)=[\bar{R}^k_m(t)]$. 
		\item \textbf{Action:} The action space is defined as all possible placement and scheduling policies for each VNF of an
		incoming service at decision TS $t$, $u_{\acute{n}n}^{fkm}(t)$ and $v_{n}^{fkm}$, the transmit power and subcarrier allocation for UL/DL access networks at decision TS $t$, $\rho^{hk}_{mn}$ and $p^{hk}_{mn}$. Let $\mathcal{A}$ \textcolor{black}{denote} the system action space, the action executed in the TS $t$, $a(t)\in\mathcal{A}$ is a vector defined as $a(t)=\left\lbrace \textbf{u}(t), \textbf{v}(t),\boldsymbol{\rho}(t),\textbf{p}(t)\right\rbrace$,
		where $\textbf{u}(t)=[u_{\acute{n}n}^{fkm}(t)],\textbf{v}(t)=[v_{n}^{fkm}(t)],\boldsymbol{\rho}(t)=[\rho^{hk}_{mn}(t)],\textbf{p}(t)=[p^{hk}_{mn}(t)]$.
		\item \textbf{Reward:}
		Since our objective in this paper is to jointly minimize the network cost and average AoI of the destination users, so we define the reward function as negative sum of total network cost and average AoI as follows:		
		\begin{align}\label{eq-reward}
	R(t)=\begin{cases}
		-\Theta_3, & \text{if (\ref{eq-u-1})-(\ref{eq-3}) are satisfied.}\\
		-\infty, & \text{otherwise}.
	\end{cases} 
	\end{align}
	\end{itemize}
	
	Where $\Theta_3={\xi\bar{\Delta}+\xi_1\Lambda_1+\xi_2\Lambda_2+\xi_3\Lambda_3+\xi_4\Lambda_4}$. The negative infinity in reward function used in order to tell the agent how much is important to choose actions in order to satisfy constraints (1) to (\ref{eq-3}). In other words, if agent cannot satisfy the mentioned constraints, it will be punished with very big negative reward.
	To handle the discrete VNF placement and subcarrier allocation with the continuous transmit power allocation, the optimal resource allocation selection policy is divided into two parts, i.e., a policy for joint selecting the optimal node and proper subcarriers (discrete actions), and  a policy for
	selecting the optimal transmit power allocation (continuous actions) \cite{hu20201}. The optimal selected node and subcarriers, which maximizes the Q-function value for the current state $s$ is determined as
	
	\begin{equation}\label{eq-231}
		a(t)=\arg\max_{a(t)}Q^*(s(t),a(t);\boldsymbol{\theta};\pi^*(a(t)|s(t);\boldsymbol{\omega})),
	\end{equation}
	where policy $\pi^*(a(t)|s(t);\boldsymbol{\omega})$ provides the optimal transmit power given state $s(t)$, selected node and subcarriers. The exact estimation of policy functions, $Q^*(s(t),a(t);\boldsymbol{\theta})$ and $\pi^*(a(t)|s(t);\boldsymbol{\omega})$ is time-consuming due to the high dimensional state and action spaces. Therefore, neural networks are adopted to approximate policy functions $Q^*$ and $\pi^*$. Moreover, the training of the neural networks is done by combining the training methods used in the DQN and DDPG algorithms. The DDPG algorithm is one of the policy-based RL algorithms and is more compatible for large and continuous state and action spaces. Under actor-critic architecture, this algorithm utilizes DNNs as function approximators for finding deterministic policies that can map large discrete or continuous states into continuous actions.\vspace{-0.1cm}
	\subsection{Proposed Multi-Agent Compound Action Actor-Critic Algorithm} 
\begin{figure}[h]
	\begin{center}
		\includegraphics[width=3.3 in]{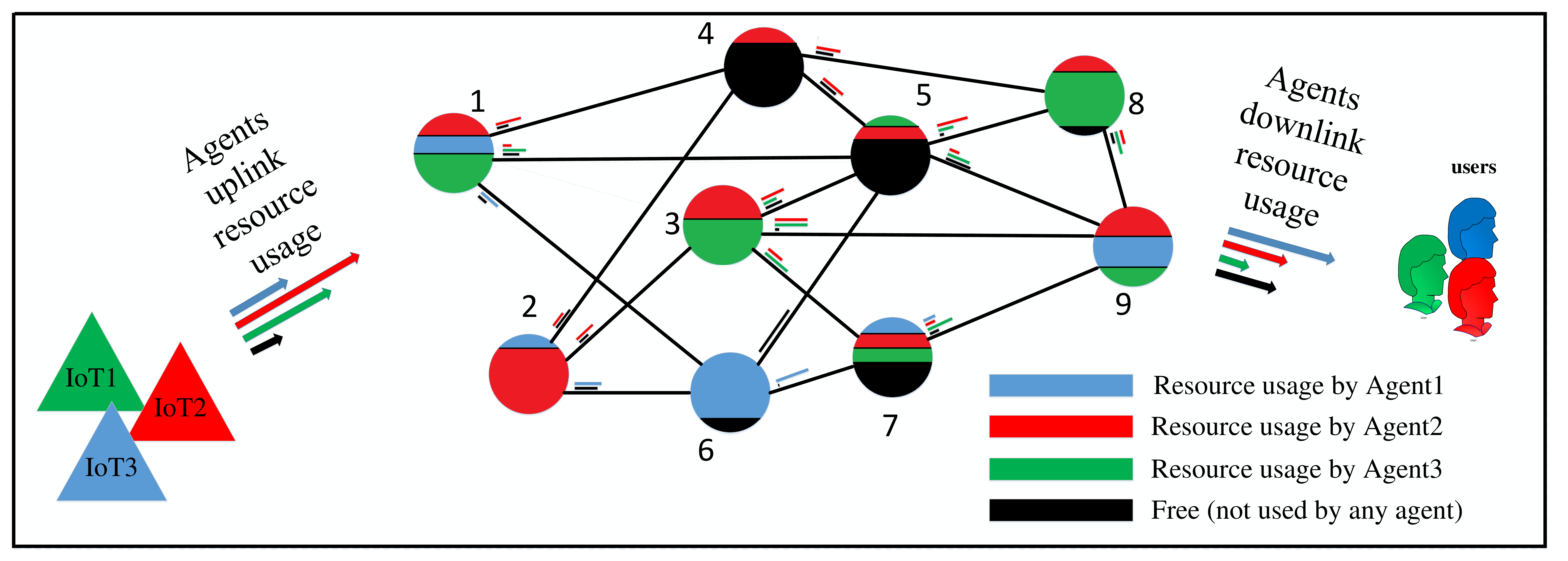} 
		\caption{The proposed MA scenario. The colors show the resource usage by each agent.} 
		\label{scenario-MA}
	\end{center}\vspace{-0.3cm}
\end{figure}

Although the centralized single-agent (SA) methods can provide better performance than distributed multi-agent (MA) methods, but they lead to large signaling overhead specially in the environment with large state space. However, in some environments like the proposed one in this paper, using multiple copies of the same agents while cooperating with each other to solve the shared problem may lead to get better results since each SA has limited capacity to learn \cite{bowling2003multiagent, panait2005cooperative}. In this section, we extend our solution to distributed method in which there are $N$ agents which must cooperate with each other to serve the arrived requested services and obtain maximum cumulative shared reward. More precisely, we have multiple agents which have access to all resources in the environment, i.e., each agent has options to consume all network resources to serve it's assigned service requests. Their shared goal is serving as maximum number of services as possible while satisfying their quality of service requirements and minimizing joint network cost and average AoI of destination users. Fig. \ref{scenario-MA} shows an illustrating example of the proposed MA scheme. As depicted in Fig. \ref{scenario-MA}, three agents consume the network resources in order to serve the service requests. Each service request is assigned an agent with the same index, i.e., IoT1, IoT2 and IoT3 services are assigned to agent1, agent2 and agent3, respectively. The colors are indicating that how much resource is used by each agent. For example, a few amount of resource from the node number 4 in the Fig. \ref{scenario-MA} is just utilized by agent2 (red color) and has high residual free resource (black color). It should be mentioned that sometimes the agents may exceed the network resources in order to serve the services and consequently, get negative rewards. Let $\mathcal{S}_n(t)$ \textcolor{black}{denote} the n$th$ agent local
	observation defined as $\mathcal{S}_n(t)=\left\lbrace \textbf{r}^{\text{BW}}_n(t), \textbf{r}^{\text{M}}_n(t), \textbf{r}^{\text{C}}_n(t),\textbf{g}_n(t),\bar{\textbf{R}}_n(t)\right\rbrace$, and $\mathcal{A}_n(t)$ \textcolor{black}{denotes} the n$th$ agent local
	action at TS $t$ given by $\mathcal{A}_n(t)=\left\lbrace \textbf{u}_n(t), \textbf{v}_n(t),\boldsymbol{\rho}_n(t),\textbf{p}_n(t)\right\rbrace$. Each agent interacts with the environment and takes its own action $\mathcal{A}_n(t)$ based on its own observation $s_n(t)\in\mathcal{S}_n(t)$ independent of the other agents and gets a reward $r_n(t)$.
	Therefore, in the proposed approach, agents do not need global information which significantly reduces the signaling overhead. It should be noted that each agent is considered as compound action actor-critic (CA2C) that is defined in previous section. The proposed MA scheme is based on centralized training and distributed execution. In the training phase, each agent sends its own action and observation to the critic network, and then the critic network uses all states including other agent's observations and actions for its own training. However, in the execution phase, each agent executes its action according only receiving its own observation in order to maximize the accumulated reward. To prevent the network from getting stuck in a bad local optimum trap near the initial point, we use ornstein-uhlenbeck process to add noise $\mathcal{M}_n(t)$ to the selected action to ensure that all actions are explored \cite{hou2017novel}. The details of the MA-CA2C algorithm is described in Algorithm \ref{alg1}.
	\begin{algorithm} 
		\caption{MA-CA2C Algorithm}\label{alg1}
		\small
		\begin{algorithmic}[1]
			\State  Initialize critic network $Q_n(s_n(t),a_n(t);\boldsymbol{\theta}_n)$ and actor network $\pi_n(a_n(t)|s_n(t);\boldsymbol{\omega}_n)$
			\State Initialize target networks $\acute{Q}_n(s_n(t+1),a_n(t+1);\acute{\boldsymbol{\theta}}_n)$, and  $\acute{\pi}_n(a_n(t+1)|s_n(t+1);\acute{\boldsymbol{\omega}}_n)$
			\State Initialize reply buffer length with $D$
			\For{episode from 1 to number of episodes}
			\State Receive initial state $\textbf{s}(t)$
			\For{Cycle from 1 to number of cycles}
			\For{from 1 to the required samples}
			\State Based on the current policy, select an action	for agent \\~~~~~~~~~~~~~$n$ by $a_n(t) =\pi_n(a_n(t)|s_n(t);\boldsymbol{\omega}_n)+ \mathcal{M}_n(t)$ 
			\State Execute the action and observe the reward $r_n(t)$ by (\ref{eq-reward}) \\~~~~~~~~~~~~~and the next state $s_n(t+1)$
			\If{all the optimization problem constraints are satisfied \\~~~~~~~~~~~~~in states $s_1(t+1),\dots, s_N(t+1)$}\parState{save transition $\textbf{s}(t), \textbf{a}(t), \textbf{s}(t+1), r_1(t),\dots, r_N(t)$ in $D$}
			\EndIf
			\EndFor
			\For{from 1 to number of train steps}
			\State {Sample a random minibatch of $M$ transitions from \\~~~~~~~~~~~~~replay buffer $D$}
			\State Update critic network by minimizing the loss function
			\State {Update actor network by the sampled policy gradient
			}
			\State Update the target networks.
			\EndFor
			\EndFor 
			\EndFor 
		\end{algorithmic}
	\end{algorithm}
	\section{Computational Complexity and Convergence Analysis}\label{ComputationalComplexity}
	The computational
	complexity of our proposed algorithm consists of two main parts, i.e, the computational complexity of action selection and the computational complexity of training process. \textcolor{black}{The training processes of the DLR methods are done in the offline mode, which is a widely used approach in systems that requires low response time from learning algorithms. Similar to the most recent papers in this era \cite{8876694,shah2020multi,li2018deep}, all the results and values in the figures of this paper are the ones that the agents converged after several training episodes.}
	\subsection{Computational Complexity of Action Selection}
	We assume that our neural network is a fully connected neural network with fixed numbers of hidden layers and fixed numbers of neurons in each hidden layer. The computational complexity of calculating the output of such neural network for given an input is equal to the sum of the sizes of input and output \cite{sipper1993serial}. For our proposed algorithm, based on the states and actions defined, for each agent, the sizes of the inputs of the critic and actor networks are $N+1+M+MK$ and $(N-1)FKM+FKM+2HKM$, respectively. Thus, the computational complexity of action selection and estimation of the Q-function value for a state-action pair is $\mathcal{O}(NFKM)$. The estimation of the Q-function values should be done at all $N$ agents, thus, the computational complexity of action selection is $\mathcal{O}(N^2FKM)$. 
	\subsection{Complexity of Training Process} 
	In accordance to (\ref{eq-231}), the Q-function values of the K services should be calculated and compared by the agents before the training step. Based on previous section, the computational complexity of this step is $\mathcal{O}(MBRK)$ where $M$ is the size of the training batch. In addition, for a fully connected neural network in which  the number of hidden layers and neurons are fixed, the back-propagation algorithm complexity is related to the product
	of the input size and the output size. For each node, the sizes of the inputs of the critic and actor networks are $RK+K$ and $2RK$, respectively. Moreover, for each agent, the sizes of the outputs of the critic and actor networks are 1 and 2, respectively. Thus, the back-propagation algorithm complexity is $\mathcal{O}(MBRK)$. Finally, the training process complexity is $\mathcal{O}(MBRK)$. 
	\subsection{Convergence Analysis}\label{Convergence Analysis}
	Our proposed algorithm is an extended version of Q-learning algorithm. For Q-learning algorithm, if $\sum_{t=0}^{\infty}\alpha^t=\infty$ and $\sum_{t=0}^{\infty}(\alpha^t)^2<\infty$ are satisfied and $|r^t(s^t,a^t)|$ be bounded, the Q-function converges to the optimal Q-function as $t\rightarrow\infty$ with probability 1 \cite{watkins1992q}. An effective approach to train neural networks is using of the inverse time decaying learning rate in which using of the large learning rate in the first training epochs prevents the network from getting stuck in a bad local optimum trap near the initial point. Whereas, using the small learning rate in the last training epochs converge the network to a good local optimum and prevents the network from oscillation. We also analyze the convergence of our proposed algorithm through simulations in Section \ref{SimulationResults}.
	\section{Simulation Results and Discussion}\label{SimulationResults}
\begin{figure}[h]
		\begin{center}
			\subfigure[]{\label{Reward}
				\includegraphics[width=3.1 in]
				{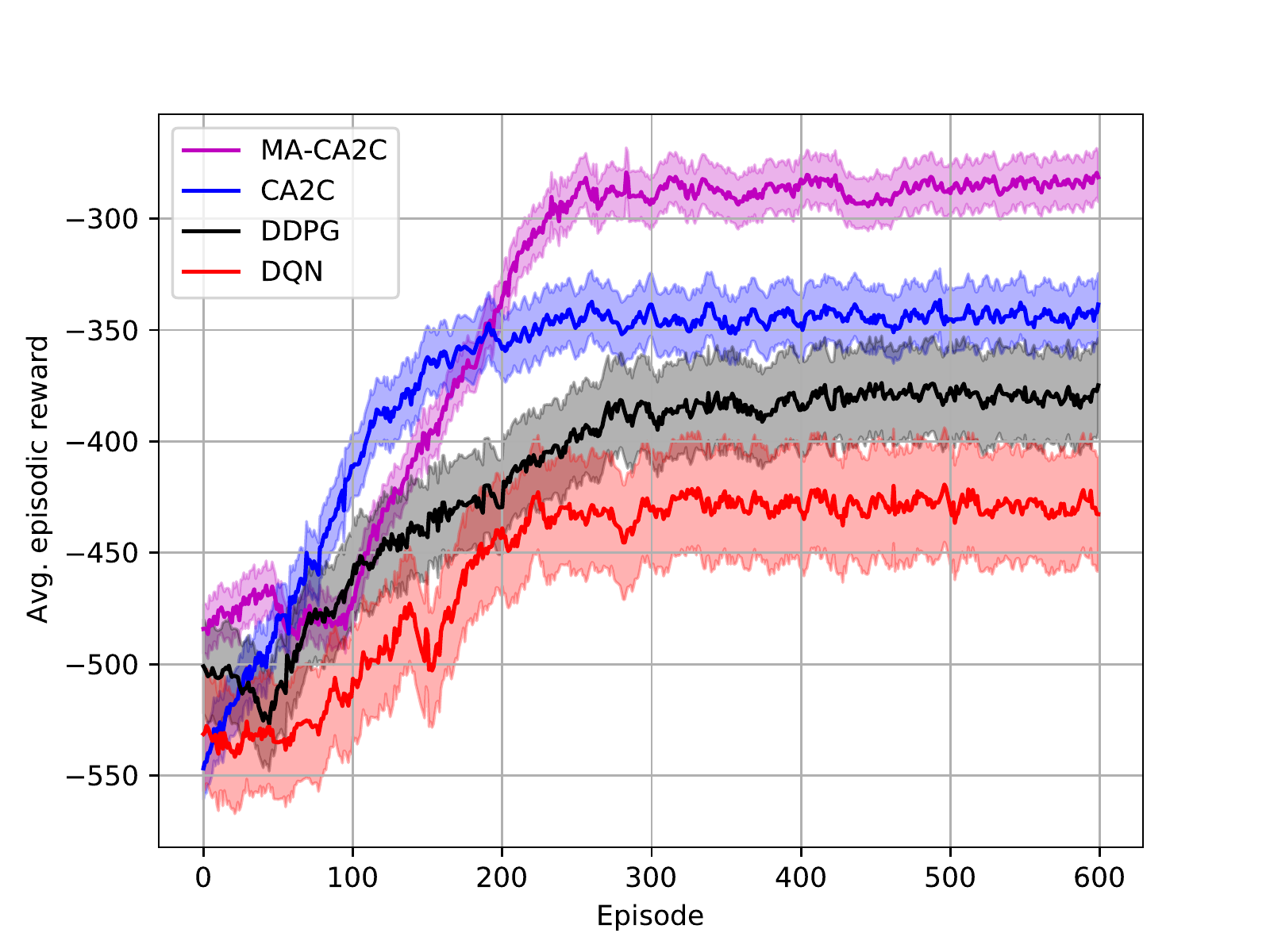}}
			\subfigure[]{\label{AoI}
				\includegraphics[width=3.1 in]
				{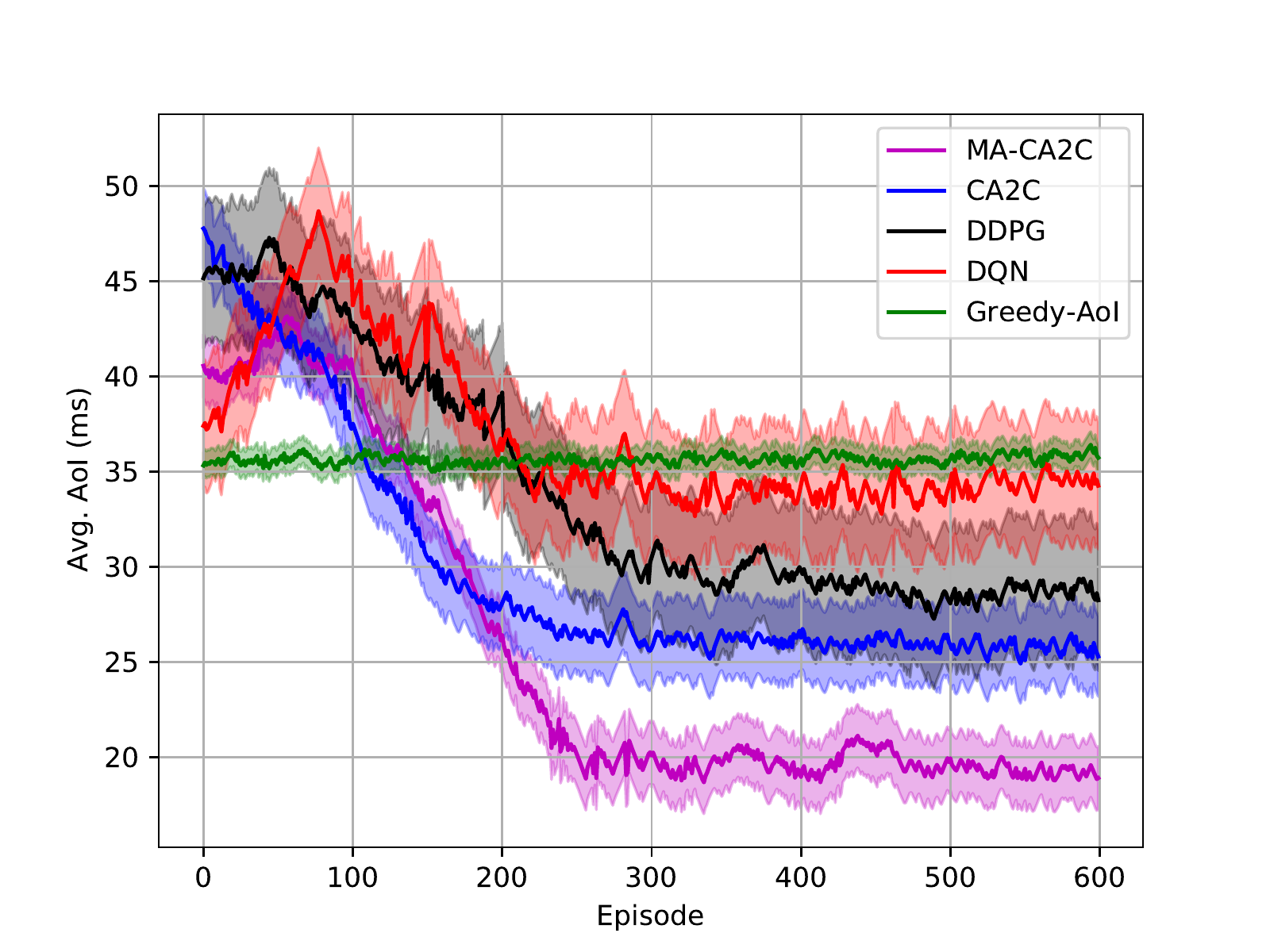}}
			\caption{(a) Average episodic reward for our proposed model with different RL methods. (b) Average AoI for the proposed model with different RL methods compare to baseline Greedy.}
		\end{center}
	\end{figure}

\begin{figure}[h]
		\begin{center}
			\includegraphics[width=3 in]{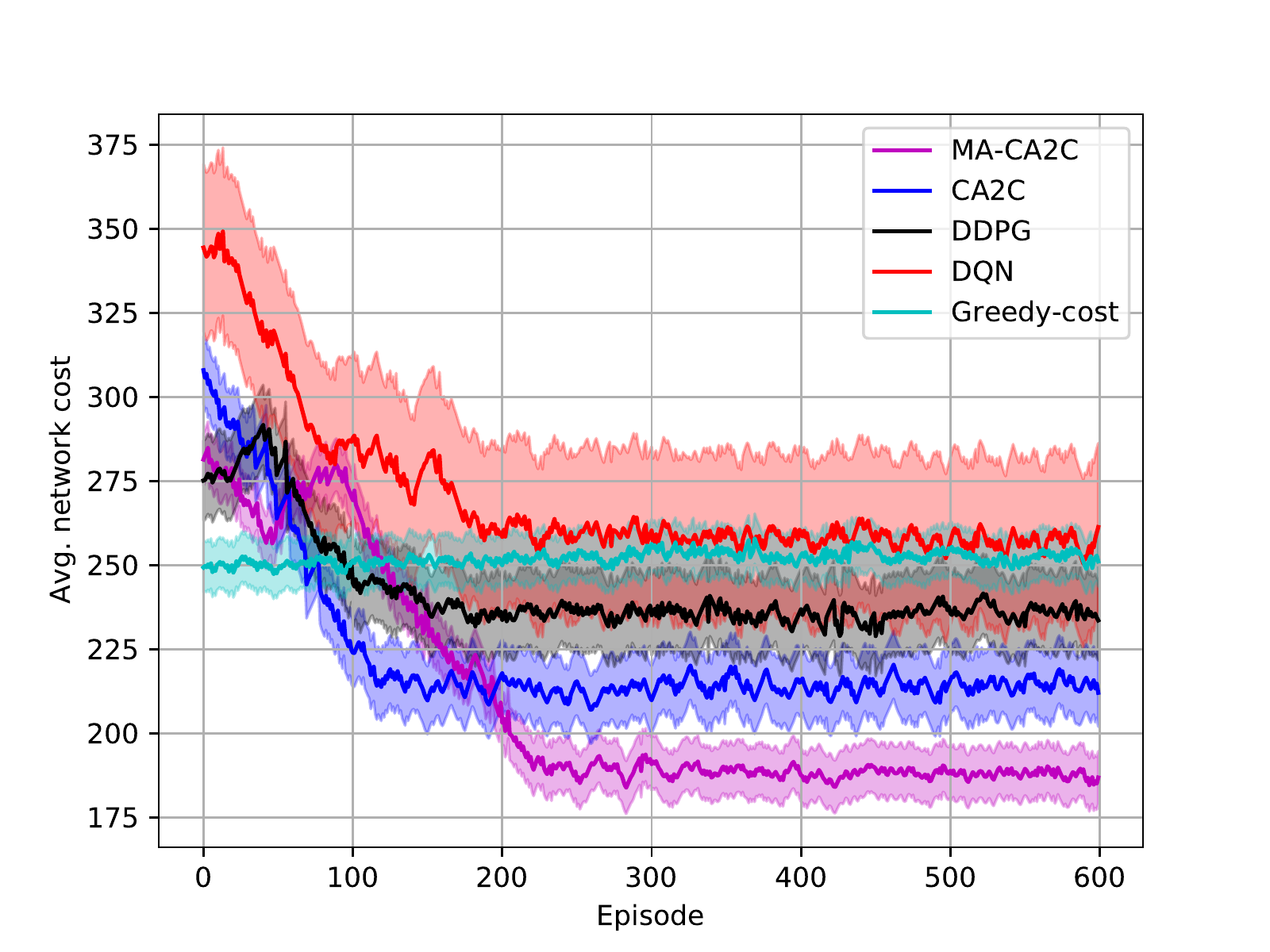} 
			\caption{Average cost for our proposed model with different RL methods compare to baseline Greedy.} 
			\label{Cost}
		\end{center}
	\end{figure}

\begin{figure}[h]
		\begin{center}
			\subfigure[]{\label{image007}
				\includegraphics[width=3.1 in]
				{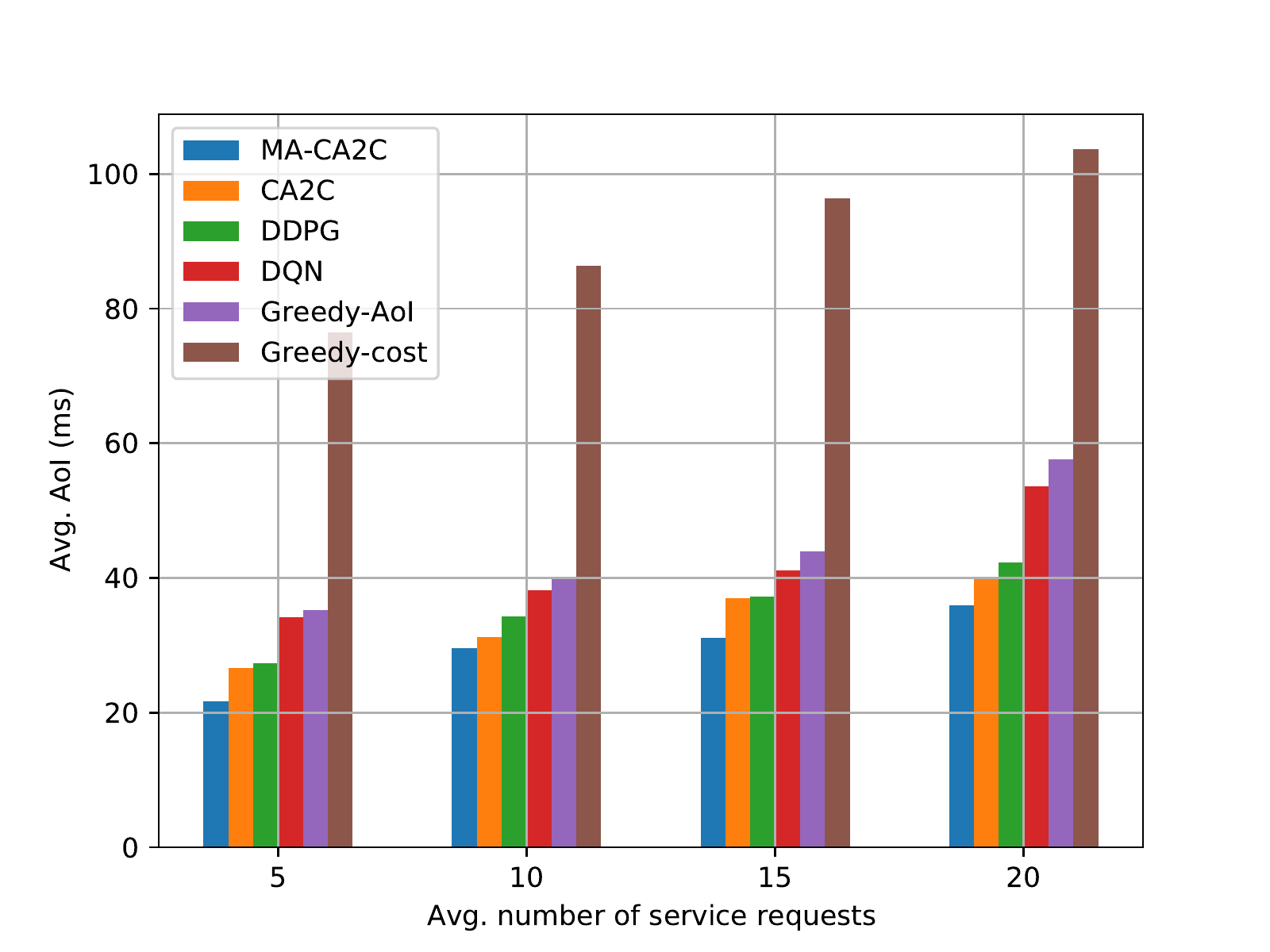}}
			\subfigure[]{\label{image009}
				\includegraphics[width=3.1 in]
				{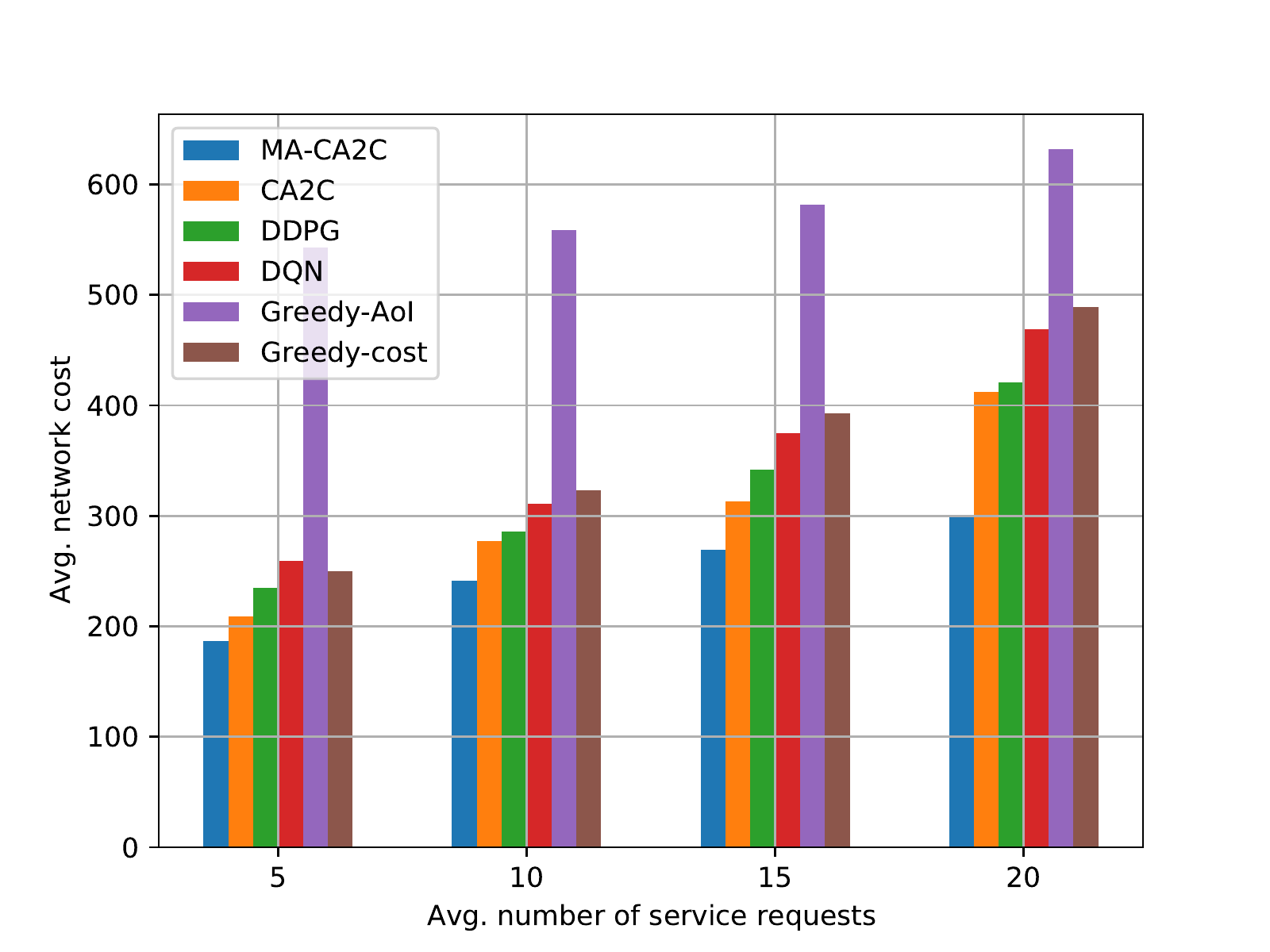}}
			\caption{(a) Average AoI VS. The number of services arrived per TS with 25 virtual nodes and 135 links. (b) Average network cost VS. The number of services arrived per TS with 25 virtual nodes and 135 links.}
		\end{center}
	\end{figure}

\begin{figure}[h!]
		\begin{center}
			\subfigure[]{\label{image011}
				\includegraphics[width=3.1 in]
				{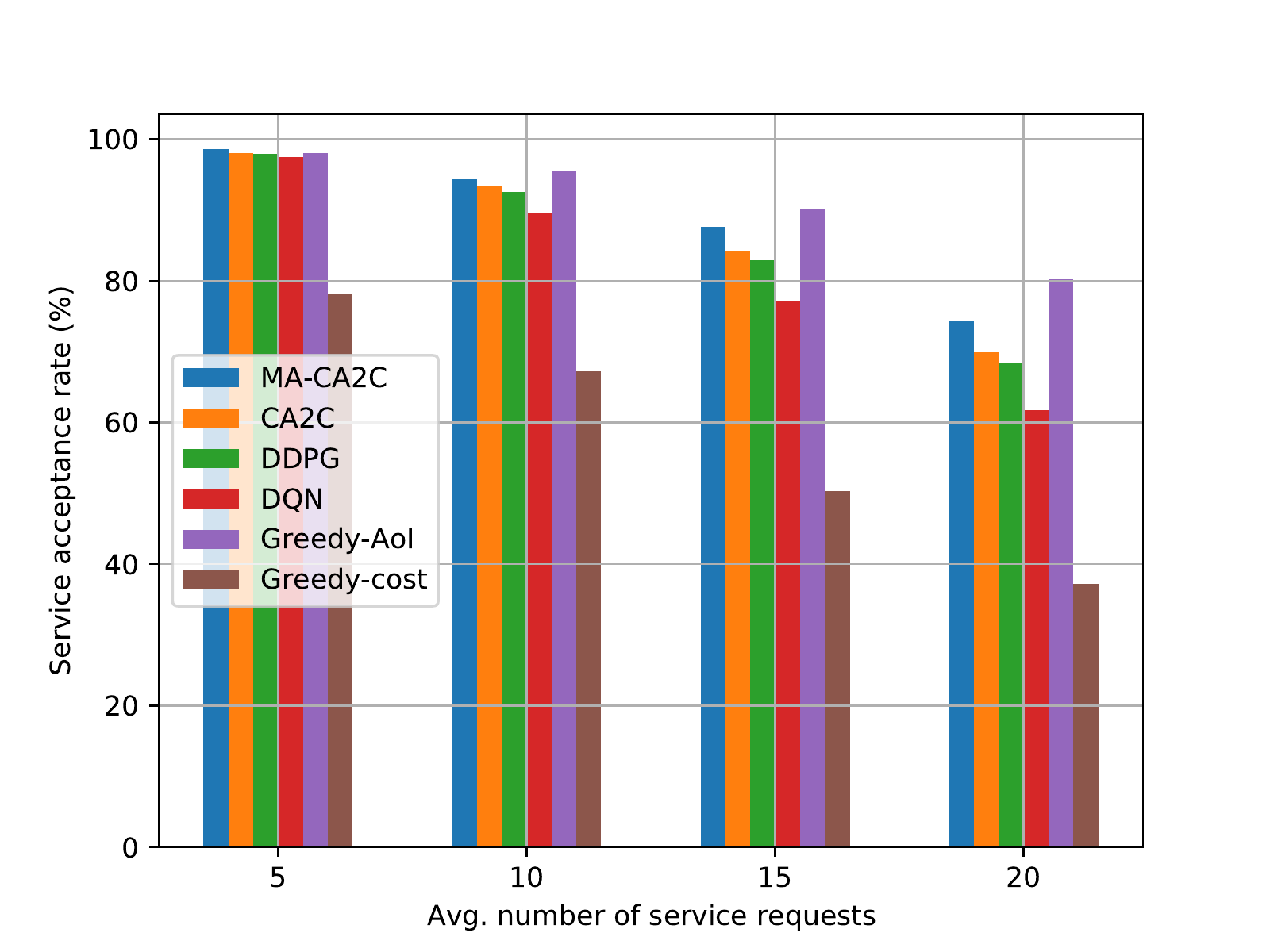}}
			\subfigure[]{\label{image012}
				\includegraphics[width=3.1 in]
				{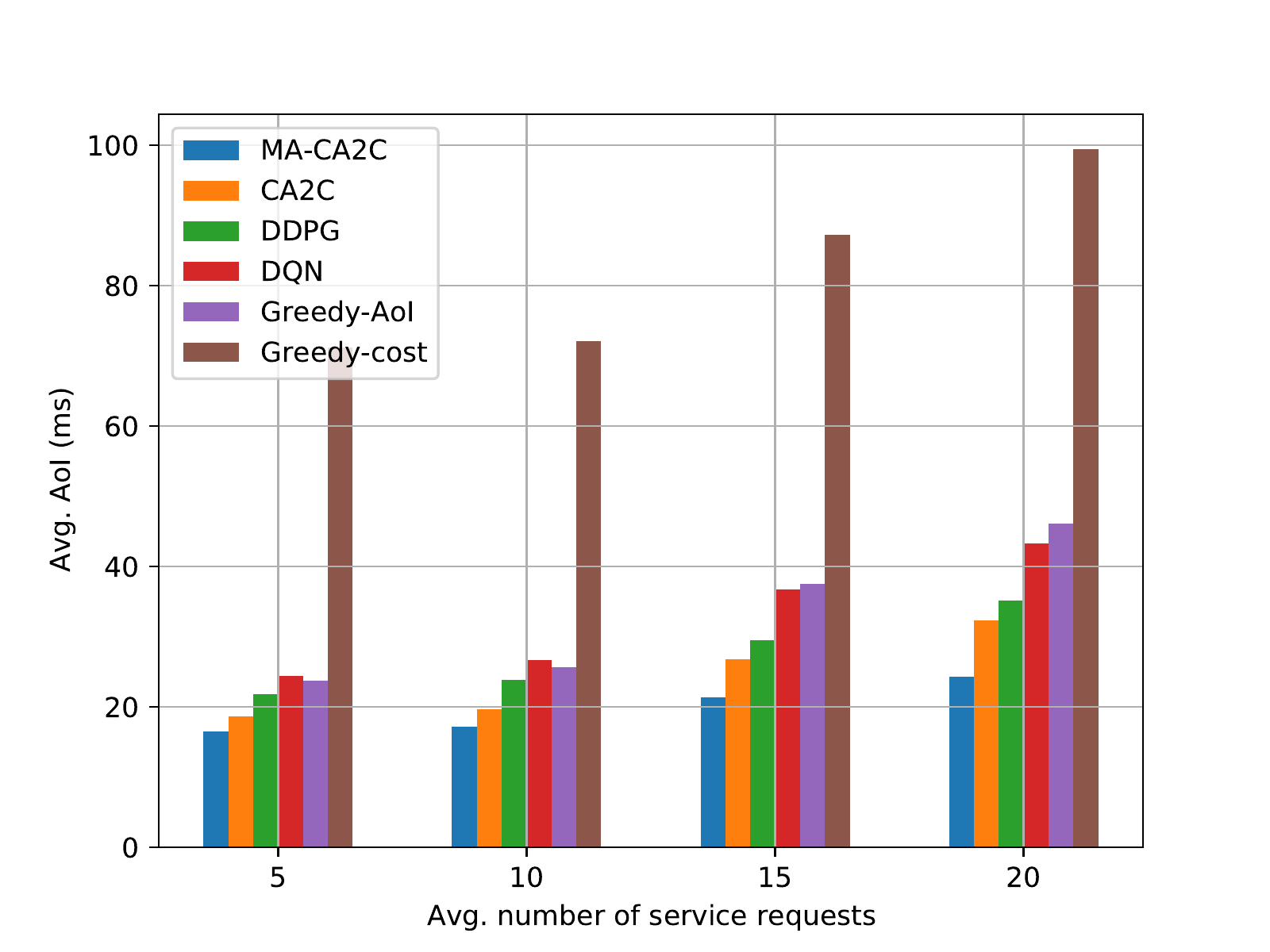}}
			\caption{(a) Acceptance rate VS. The number of services arrived per TS with 25 virtual nodes and 135 links. (b) Average AoI VS. The number of services arrived per TS with 50 virtual nodes and 270 links.}
		\end{center}
	\end{figure}

\begin{figure}[h!]
		\begin{center}
			\subfigure[]{\label{image013}
				\includegraphics[width=3.1 in]
				{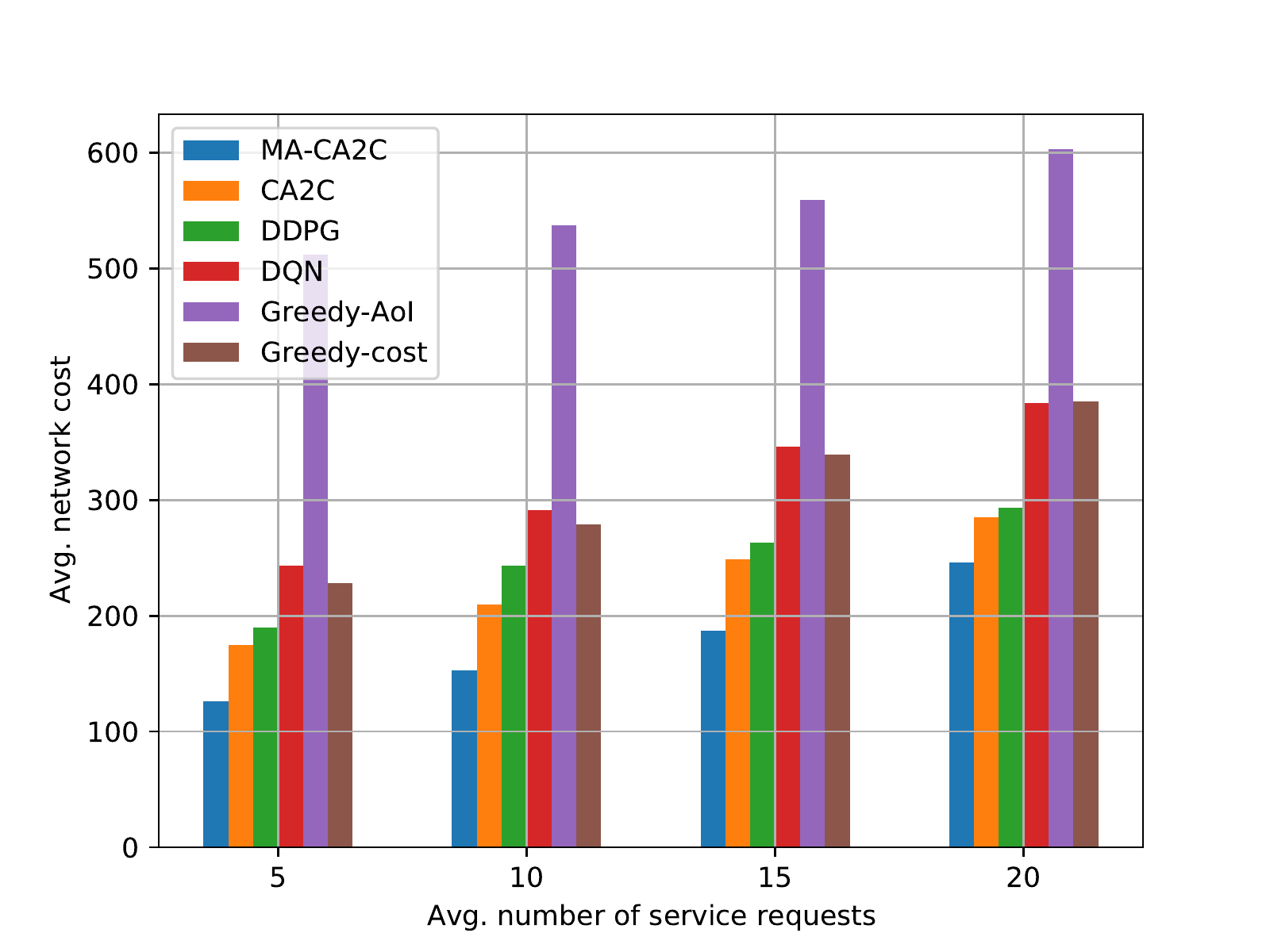}}
			\subfigure[]{\label{image014}
				\includegraphics[width=3.1 in]
				{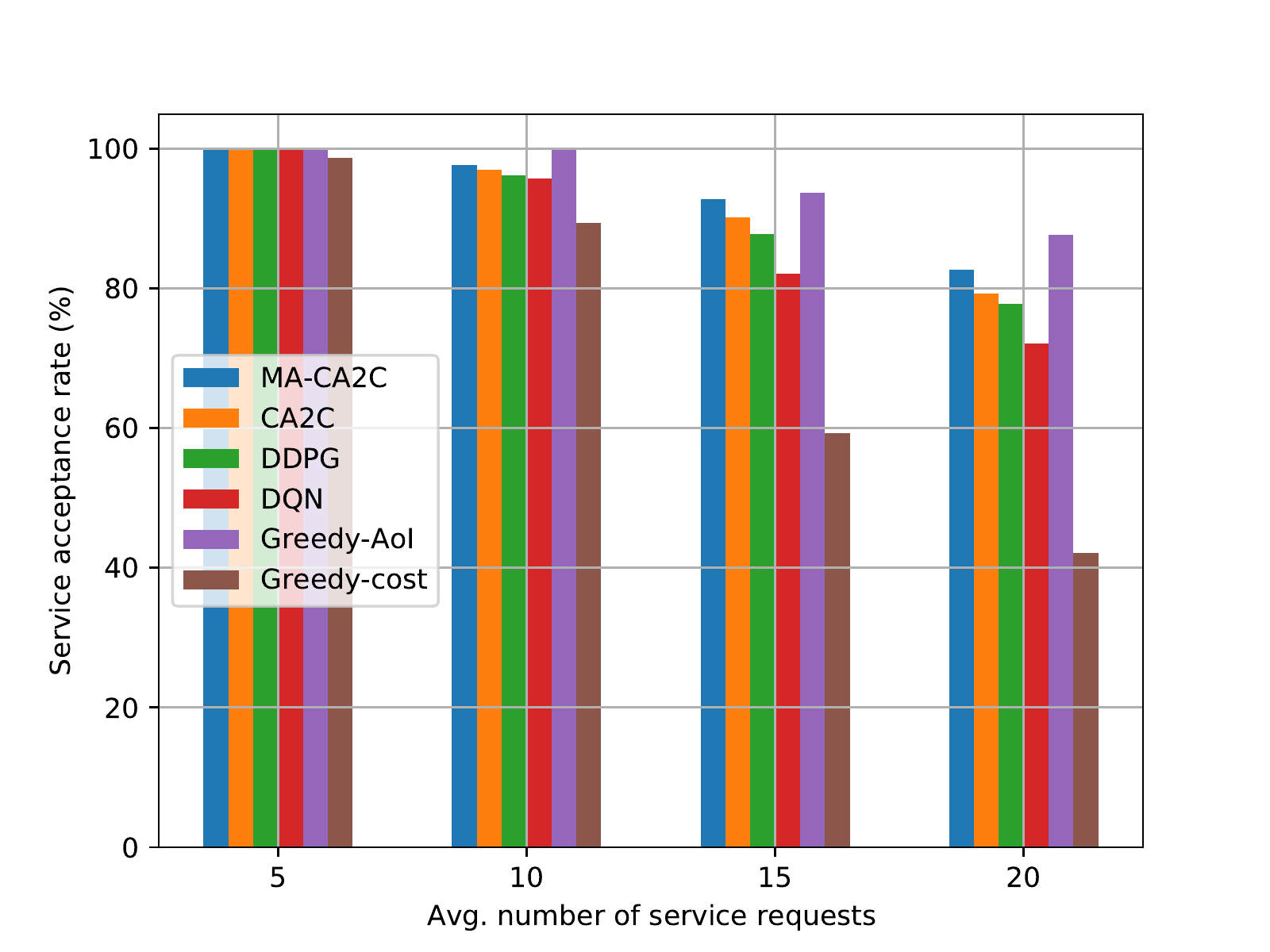}}
			\caption{(a) Average cost VS. The number of services arrived per TS with 50 virtual nodes and 270 links. (b) Acceptance rate VS. The number of services arrived per TS with 50 virtual nodes and 270 links.}
		\end{center}
	\end{figure}

	In this section, the performance of our proposed algorithm is evaluated and compared with two Greedy algorithms; Greedy-AoI and Greedy-cost. In Greedy-AoI algorithm, the paths consisting of nodes and links, are sorted in ascending order based on the latency that they can add to the services,  and the arriving service requests are placed on the paths which have lowest latency. Also, the packets of the services are sorted based on their AoI in descending order and the service with maximum AoI will be scheduled with highest priority. In Greedy-cost algorithm, the paths consisting of nodes and links are sorted in ascending order based on the average network cost that they can impose to the network. The paths that result in lowest network cost will be used for function placement. 
	In the simulation, the impact of agent cooperation, the number of nodes and the number of arrival services are evaluated on the performance of the proposed model in terms of both cost and AoI minimization. For performance
	evaluation, four different topologies consisting of 25, 50, 75 and 100 virtual nodes and 135, 270, 405 and 540 links are created and IIoT devices are uniformly distributed in 1000$\times$1000 $m^2$ area. The other simulation parameters are summarized in Table \ref{TAB-1}.
	\begin{table*}[h]
		\small 
		\setlength{\tabcolsep}{5pt}
		\caption{Network Parameters}
		\label{TAB-1} 
		\centering          
		\begin{tabular}{|l  | l|l|} 
			\hline     
			Parameter & Description& Value\\
			\hline
			$M$&\text{Number of IIoT devices} & 5\\
			\hline
			$N$&\text{Number of source/midle/destination vitual nodes} & 5/15/5, 10/30/10, 15/45/15 ,20/60/20\\
			\hline
			$L$&\text{Number of virtual links} & 135, 270, 405, 540\\
			\hline
			$C_n$&\text{The computing cpu resource in virtual node $n$} & Randomly between 1 to 2 GHz \\
			\hline
			$B_n$&\text{The storage capacity in virtual node $n$} & Randomly between 50 to 100 GB\\
			\hline
			$W_l$&\text{The available bandwidth on link $l$} & Randomly between 100 to 1000 Mbps \\
			\hline
			$K$&\text{Number of services} & 5, 10, 15, 20\\
			\hline
			$H/\check{H}$&\text{Number of subcarriers} & 10\\
			\hline
			$\bar{R}^k_m$&Minimum bit rate required for service $k$ of IIoT device $m$& 50 Mbps\\
			\hline
			$N/A$&Number of function(s) per service & Random (4 to 7)\\
			\hline
			$N/A$&Number of packet(s) per service & Random (2 to 6)\\
			\hline
			$P_{\text{max}}$& Maximum transmit power & 30 Watt \\
			\hline
			$\iota$&TS interval &0.5 s\\
			\hline
			$\alpha_c/\alpha_a$&Critic/Actor initial learning rate & 0.005/0.001\\
			\hline
			$\gamma$&Discount factor &0.99\\
			\hline
			$N/A$&Batch-size &64\\
			\hline
			$B$&Bandwidth &15 MHz\\
			\hline
			$N_0$&Noise power spectral &-170 dBm/Hz\\
			\hline
			$\kappa$&Path loss component&3.5\\
			\hline
			$\varepsilon$& Exploration rate for DQN&0.001 \\
			\hline
			$\mathcal{M}_{t}$& Exploration noise for Actor Critic&Ornstein-Uhlenbeck \\
			\hline
			$N/A$&Activation function& ReLU\\
			\hline
			$N/A$&Number of episodes& 60000\\
			\hline
			$N/A$&Target network update frequency& 1000\\
			\hline
			$N/A$&Number of hidden layers &4 \\
			\hline
			$N/A$&Number of neurons in each layer &512 \\
			\hline
		\end{tabular}%
	\end{table*}
	The average episodic reward for our proposed model considering 25 total virtual nodes, 135 total links and average 5 service request arrival rate is depicted in Fig. \ref{Reward} for different methods. As can be seen, by MA, our proposed model can achieve better episodic reward compared to the other methods. The reason is that in MA, we use multiple copies from SA to be trained in the environment and therefore we have more capacity of training. On the other hand, in early episodes, SA, i.e., CA2C has higher episodic reward and it learns faster than MA-CA2C, because in early episodes in MA, the agents need more training steps to cooperate. Also, CA2C can learn faster and obtain higher episodic reward than DDPG, because we have both continuous and discrete actions in the environment and DDPG only supports continuous actions.
	On the other hand, DQN only supports discrete actions and since we use quantization for supporting continuous actions, DQN has worst performance compared to the other SA methods. As shown in Fig. \ref{Reward}, the convergence training episode for DQN is lower than DDPG, i.e., DQN converges after 230 training episodes while DDPG converges after 290 training episodes. This is expected since in DDPG agent has infinite number of continuous actions rather than DQN that has finite number of discrete actions. The high oscillation in the episodic reward for DQN is because of quantization of the actions. Fig. \ref{AoI} and Fig. \ref{Cost} show the average AoI of the destination user and average network cost for each training episodes considering 25 total virtual nodes, 135 total links and average 5 service request arrival rate, respectively.
	From Fig. \ref{AoI} and Fig. \ref{Cost} it could be observed that the DRL methods can significantly outperform the greedy algorithms since they learn to predict the behavior of the network and incoming service requests. 
	Fig. \ref{image007} and Fig. \ref{image009} demonstrate the average AoI and network cost versus the arrival rate of service requests with total virtual nodes equals to 25. As observed from Fig. \ref{image007} and Fig. \ref{image009}, increasing the service request rate, while maintaining the total virtual nodes and links fixed to 25 and 135, compels the agent to consume more network resources. Fig. \ref{image011} shows the acceptance rate versus the number of service request with 25 virtual nodes and 135 links. Although, as depicted the Greedy-AoI algorithm had accepted more service requests than other RL methods, but it \textcolor{black}{increases} the network cost drastically as shown in Fig. \ref{image009}. It should be noted that our proposed solution jointly minimizes the AoI and the network cost, so some service requests will be rejected in order to decrease the network cost. 
	Fig. \ref{image012} and Fig. \ref{image013} show the AoI and network cost for 50 virtual nodes and 270 links, respectively. As expected, increasing the number of service request from 5 to 20 causes to increase the AoI and network cost. 
	Fig. \ref{image014} depicts the acceptance rate comparison for 50 nodes and 270 links, while the service request rate increases from 5 to 20. Since the capacity of SDN increases, the acceptance rate improves, as well. For the sake of brevity, the other simulation results for other nodes and service requests are not provided.
	\section{Conclusion and Future Work}\label{conclusions}
	In this paper, we developed single-agent and cooperative multi-agent CA2C DRL-based VNF placement and scheduling for virtualized IIoT network to minimize VNF placement cost, scheduling cost, and average AoI. Our proposed single-agent CA2C DRL-based VNF placement methods significantly outperform DDPG and DQN methods and greedy algorithms in terms of average network cost and age of information. In addition, due to the capacity limitation in the single agent scheme, proposed cooperative multi-agent DRL is able to achieve higher reward. As our future work, we plan to extend our proposed solutions to network slicing where each type of user belongs to one slice. Network slicing structure allows the creation of multiple virtual networks with different rate and latency requirements atop a shared physical infrastructure.
	
	\hyphenation{op-tical net-works semi-conduc-tor}
	\bibliographystyle{IEEEtran}
	\bibliography{IEEEabrv,Bibliography}
	\vspace{-0.5 cm}
	\begin{biography}[{\includegraphics[width=1in,height=1.5in,clip,keepaspectratio]{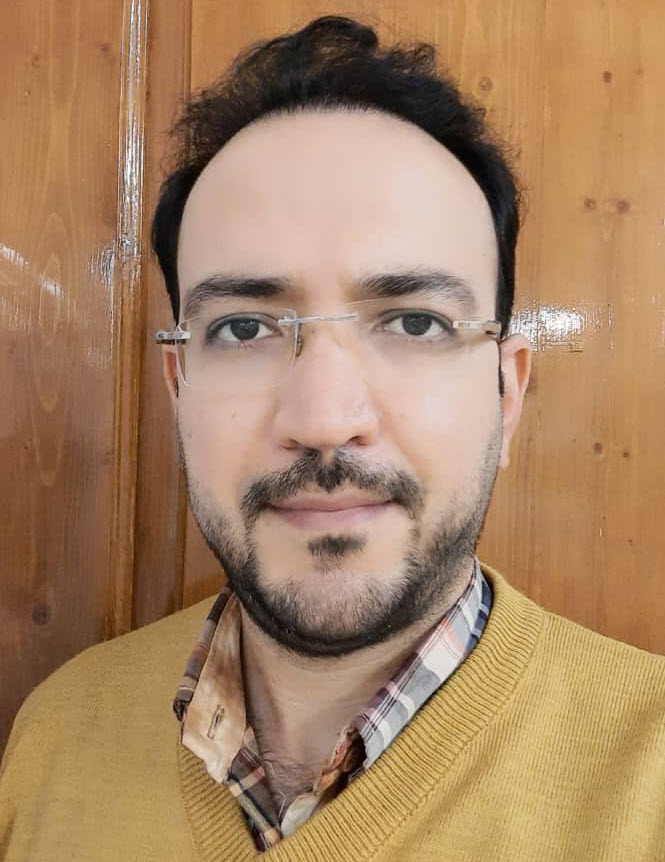}}]{Mohammad Akbari} received his B.Sc. in Electrical Engineering in 2008 from Tabriz University, Tabriz, Iran and the M.Sc. and Ph.D. degrees both from Iran University of Science and Technology (IUST), Tehran, Iran in 2010 and 2016 respectively. During 2010-2017, he was a senior system designer at Afratab R\&D group, Tehran, Iran. In 2017, he joined as a research assistant professor to the Department of Communication Technology, ICT Research Institute (ITRC), Tehran, Iran. His current research interests span topics in telecommunication system and networks including Self-Organizing Networks, 5G and 6G Networks and application of Machine Learning techniques in wireless communication.  
		\end{biography}	

	\begin{biography}[{\includegraphics[width=1in,height=1.5in,clip,keepaspectratio]{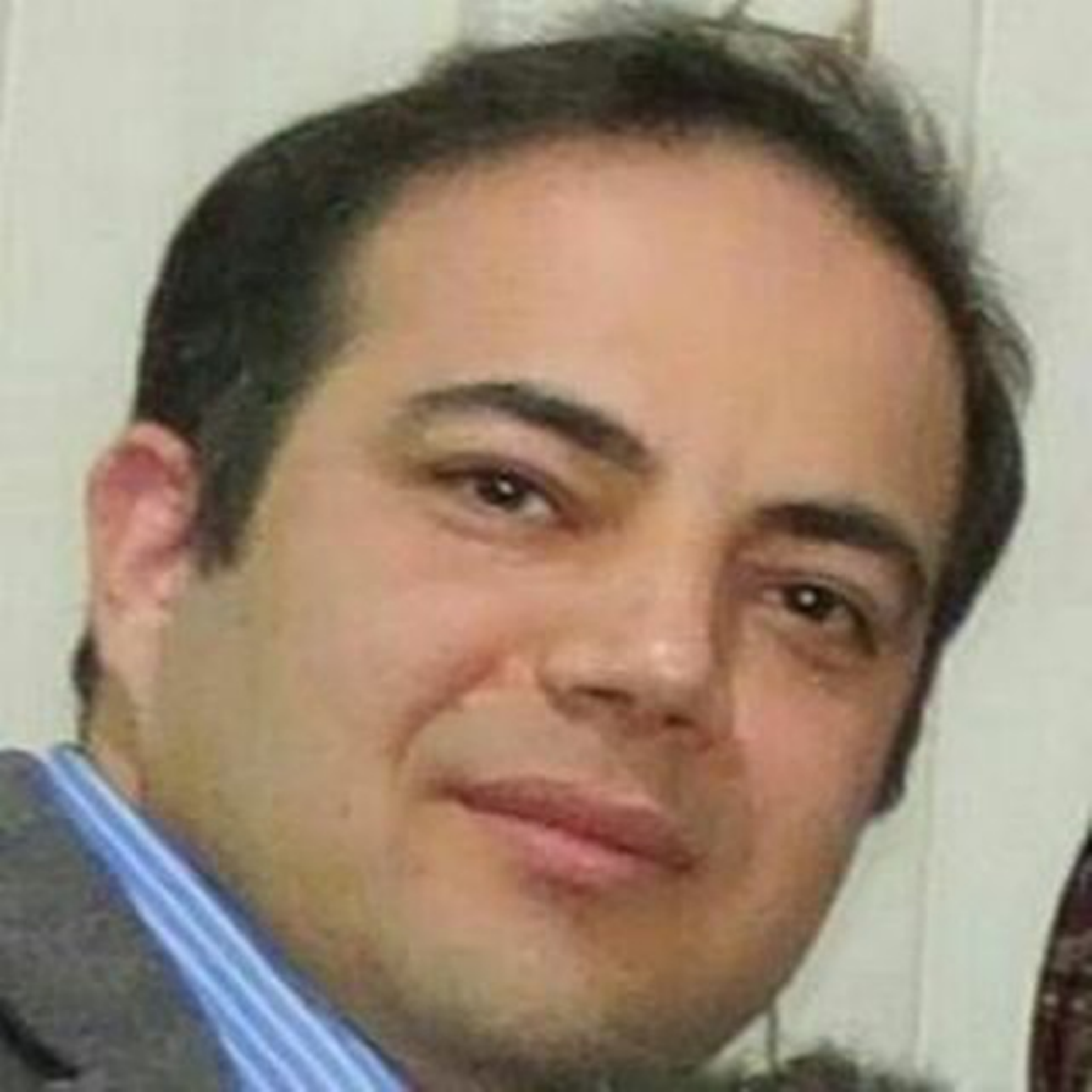}}]{Mohammad Reza Abedi} received the M.Sc. degree in electrical engineering from AmirKabir University, Tehran, Iran. He is currently working as a Research Assistant and phD student with Tarbiat Modares University, Tehran. He was a Reviewer for several IEEE journals such as, the IEEE TRANSACTIONS ON SIGNAL PROCESSING and IEEE TRANSACTIONS ON WIRELESS COMMUNICATIONS. He has been involved in a number of large scale network design and consulting projects in the telecom industry as a principle investigator or consultant.  He was also a Member of Technical Program Committees for the IEEE Conferences. His research interests include multiple access techniques, energy harvesting and wireless power transfer, cooperative and adaptive wireless communications, wireless edge caching, mobile edge computing, multibitrate video transcoding, software defined networking, wireless network virtualization, and optimization theory.
	\end{biography}	
	\begin{biography}[{\includegraphics[width=1in,height=1.5in,clip,keepaspectratio]{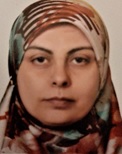}}]{Roghayeh Joda} (M’14) received the B.Sc. degree in electrical engineering from Sharif University of Technology, Tehran, Iran, in 1998, and the M.Sc. and the Ph.D. degrees in electrical engineering from University of Tehran, Tehran, in 2001 and 2012, respectively. She was a Post-Doctoral Fellow at University of Padua, Padua, Italy from September 2013 to August 2014. In November 2014, she joined ICT research institute, Tehran, Iran as a research assistant professor where she was the manager of two mega projects on 5G networks. She is currently a visiting researcher at University of Ottawa, Ottawa, Canada. Her current research interests include communication theory, information theory, resource allocation, optimization and machine learning with application to wireless networks, 5G and 6G networks
	\end{biography}	
	\begin{biography}[{\includegraphics[width=1in,height=1.5in,clip,keepaspectratio]{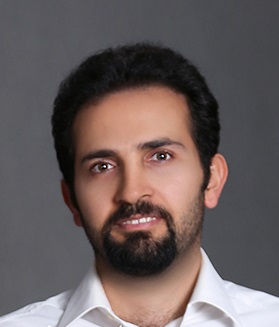}}]{Mohsen Pourghasemian} received the B.Sc. degree in electrical engineering from Bu-Ali Sina University, Hamedan, Iran, and the M.Sc. degree in electrical engineering from IRIB University, Tehran, Iran. From August 2016 to June 2018, he has participated in 5G IoT projects as a consultant at ICT Research Institute. He has been a lecturer at IRIB University from September 2018 to present. He is currently working as a Research Assistant at Tarbiat Modares University, Tehran, Iran. His research interests include wireless networks, machine learning, autonomous driving, Internet of Things, wireless sensor networks, and information theory.
	\end{biography}
	\begin{biography}[{\includegraphics[width=1in,height=1.5in,clip,keepaspectratio]{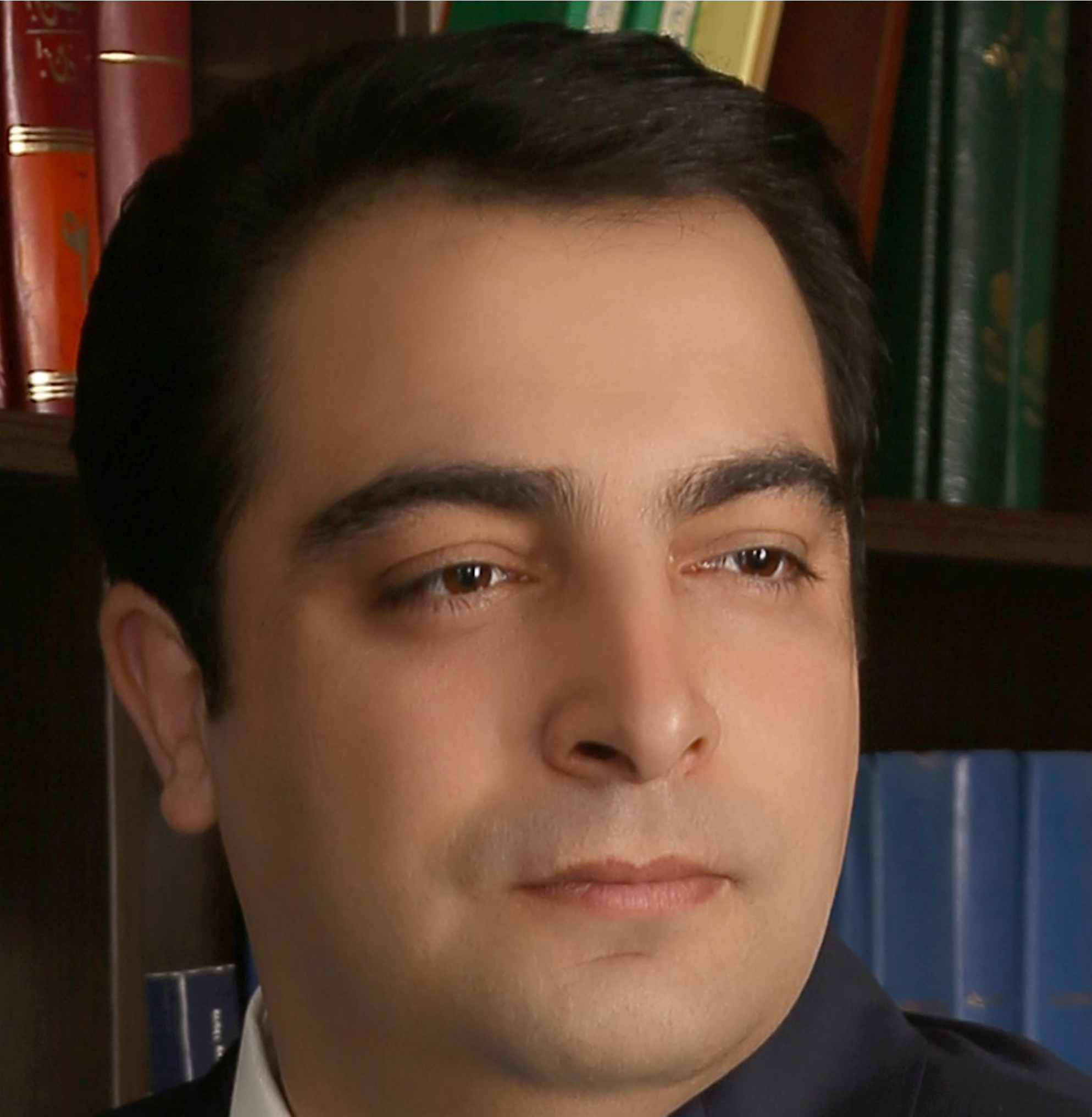}}]{Nader Mokari Yamchi} completed his PhD studies in electrical Engineering at Tarbiat Modares University, Tehran, Iran in 2014. He joined the Department of
	Electrical and Computer Engineering, Tarbiat Modares University as an assistant professor in October 2015. He was also involved in a number of large scale	network design and consulting projects in the telecom industry. His research interests include design, analysis, and optimization of communications networks.
	\end{biography}
	\begin{biography}[{\includegraphics[width=1in,height=1.5in,clip,keepaspectratio]{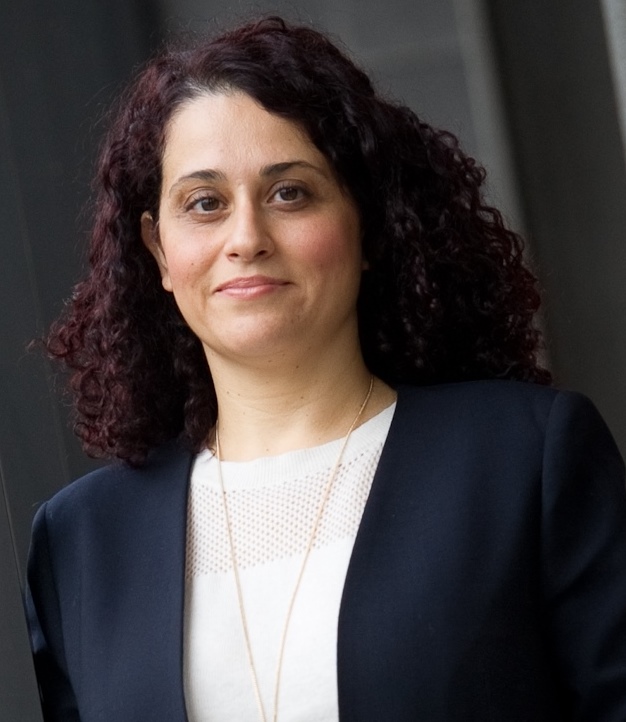}}]{Melike Erol-Kantarci} is Tier 2 Canada Research Chair in AI-enabled Next-Generation Wireless Networks and Associate Professor at the School of Electrical Engineering and Computer Science at the University of Ottawa. She is the founding director of the Networked Systems and Communications Research (NETCORE) laboratory. She is also a Faculty Affiliate at the Vector Institute, Toronto. She has received many awards and recognitions, delivered 50+ keynotes, plenary talks and tutorials around the globe. She is on the editorial board of the IEEE Transactions on Cognitive Communications and Networking, IEEE Internet of Things Journal, IEEE Communications Letters, IEEE Networking Letters, IEEE Vehicular Technology Magazine and IEEE Access. She has acted as the general chair and technical program chair for many international conferences and workshops. Her main research interests are AI-enabled wireless networks, 5G and 6G wireless communications, ORAN, smart grid, and Internet of Things. She is a senior member of the IEEE and the ACM, and a ComSoc Distinguished Lecturer.
	\end{biography}
\end{document}